\documentclass[preprint]{aastex}
\usepackage{lineno}
\linenumbers
\shorttitle{Kepler AGN}
\shortauthors{Wehrle et al.}
\begin{document}
\title{Kepler Photometry of Four  Radio-Loud AGN in 2010-2012}
\author{Ann E.~Wehrle\altaffilmark{1}, Paul J.~Wiita\altaffilmark{2}, Stephen C.~Unwin\altaffilmark{3},, Paolo Di Lorenzo\altaffilmark{2}, Mitchell Revalski\altaffilmark{2}, Daniel Silano\altaffilmark{2}, and Dan Sprague\altaffilmark{2} } 
\altaffiltext{1}{Space Science Institute, 4750 Walnut Street, Suite 205, Boulder, CO 80301}
\email{awehrle@spacescience.org}
\altaffiltext{2}{Department of Physics, The College of New Jersey, PO Box 7718, Ewing, NJ 08628, USA}
\altaffiltext{3}{Jet Propulsion Laboratory, Mail Stop 321-100, 4800 Oak Grove Drive, Pasadena, CA 91109}
\date{ Submitted to ApJ on 28 February 2013} 
                                   
\begin{abstract}
We have used Kepler photometry to characterize variability in four radio-loud active galactic nuclei (three quasars and one object tentatively identified as a Seyfert 1.5 galaxy)  on timescales from minutes to months, comparable to the light crossing time of the accretion disk  around the central supermassive black hole or the base of the relativistic jet.      Kepler's  almost continuous observations provide much better temporal coverage than is possible from ground-based observations. We report the first such data analyzed for quasars. We have constructed power spectral densities using 8 Kepler quarters of  long-cadence (30-minute) data for three AGN, 6 quarters for one AGN and 2 quarters of short-cadence (1-minute) data for all four AGN.  On timescales longer than about 0.2--0.6 day,  we find red noise with mean power-law slopes ranging from -1.8 to -1.2, consistent with the variability originating in turbulence either behind a shock or within an accretion disk.  Each AGN has a range of red noise slopes which vary slightly by month and quarter of observation.  No quasi-periodic oscillations of astrophysical origin were detected.   We detected several days-long flares when brightness increased by 3\% -- 7\% in two objects. No flares on timescales of minutes to hours were detected.   Our observations imply that the duty cycle for enhanced activity in these radio-loud AGN is small. These well-sampled AGN light curves provide an impetus to develop more detailed models of turbulence in jets and instabilities in accretion disks. 
\end{abstract}

\section{Introduction}
One of the best ways to probe the extremely small regions  from which the bulk of the energy in AGN is emitted is through the study of their variability in different bands.  
The immense powers, non-thermal spectra, and rapid variability detected across the electromagnetic spectrum that characterize the class of blazars, i.e., Flat Spectrum Radio Quasars (FSRQs) and BL Lac objects, can only be understood within the framework of matter flowing inwards through accretion disks (ADs) onto supermassive black holes (SMBHs). The bulk of this emission emerges from within several gravitational radii of the SMBH; part of the energy often is channelled outward via relativistic jets, producing radio-loud AGN.  That emission is magnified by Doppler boosting when the jet is pointed within a few degrees to our line of sight.  Such small viewing angles cause ordinary radio galaxies and radio loud quasars to appear as blazars \citep[e.g.,][]{up95}.

Optical variability can originate in relativistic jets  or in the  accretion disk \citep[e.g.,][]{mg85,mw93}.  
As the originating physical processes are different in each case, the light curves should look different. When FSRQs are in a high state, synchrotron emission from the relativistic jet overwhelms emission from the  AD.  When FSRQs are in a low state, the AD can become visible, and be recognizable as the Big Blue Bump \citep{sm89}.  During these low and high states, which can differ by 2 to 5 magnitudes for the most active blazars (optically violent variables; OVVs), the dominant light source should determine the variability characteristics of the FSRQs, specifically, the power spectral densities (PSDs) and possible existence of quasi-periodic oscillations.    

The Kepler mission \citep{borucki2010, koch2010} is uniquely able to probe the innermost regions of AGN and produce superior light curves through its ability to monitor uninterrupted by long gaps; it has only brief one-day  gaps every month for data downlink.  This capability allows us to study a broad range of variability time scales.  The overwhelming majority of {\it monitored} blazars display significant microvariability  (variations of at least 0.03 mag) on timescales less than a day 
\citep[see review by][]{miller96}; however, it is not known if this behavior is characteristic of all blazars because observers tend to select the most variable objects to monitor.  We selected radio-loud AGN in the Kepler field of view for variability monitoring observations obtained over 8  quarters in 2010-2012.  
Variability probably probes the characteristics of the accretion disks when the sources are in quiescent faint states and allows us to study the synchrotron jet emission when the sources are in highly active states.
For both radio-loud and radio-quiet quasars, there is a good probability of observing significant variability (a few tenths of a magnitude) over the course of two years \citep[e.g.][]{pica88,hawk02,macleod12}.  In a study of blazars observed with the Palomar Quest survey, \cite{bauer2009} found that 35\% of blazars showed $ V > 0.4 $ magnitudes of variation over 3.5 years.  

In addition, Kepler is capable of detecting  microvariability over the course of several hours at the level of a few percent for targets brighter than about 17th magnitude.  Ground-based optical studies have indicated that  low frequency peaked blazars and core-dominated radio galaxies with high polarizations frequently exhibit  stronger microvariability, as expected if the jets point toward us and benefit from Doppler boosting.  Ordinary radio-loud quasars and radio-quiet quiet quasars also show some microvariability, but it is both  less frequent and typically weaker 
\citep[e.g.,][]{gk03,carini07,ramirez09,goyal12}.   Core-dominated radio-loud quasars, which are believed to have jets pointing close to our line-of-sight, and therefore expected to show substantial fast variations,  exhibit much more microvariability than radio-quiet quasars if they also exhibit high optical polarizations \citep{goyal12}.  

This paper is the first report on Kepler data for FSRQs. \cite{mushotzky11} reported Kepler monitoring of 4 Seyfert galaxies ($z < 0.09$), each for a duration of 2 to 4 quarters; we note that our targets have larger redshifts and are substantially fainter (by 2 to 3 magnitudes).  
According to their analysis, all four Seyferts showed some degree of variability over these periods and all exhibited  very steep red noise components to their power spectral densities (PSDs; $-2.6 \ge \alpha \ge -3.3$).   \cite{cr12} provided a more in-depth analysis of the Kepler data for one of those Seyferts, II Zw 229.015, and were able to combine Kepler data across 3 quarters because they had sufficient ground-based measurements to  normalize the fluxes on different detectors of Kepler's camera during different quarters.  This allowed them to extend the PSD to lower frequencies where they detected a flattening below a frequency corresponding to $\sim$44 days; their best fit to the data in the regime measured by \cite{mushotzky11} led them to a shallower PSD slope of around $-2.8$.   Our results on FSRQs show significantly flatter PSDs than either group; we discuss this in \S 6.

In this paper, we first briefly review in \S 2 the physical origins of the variability signatures that disks and jets can be expected to exhibit, then describe our target sample selection in \S 3. In \S 4 we explain the steps needed to reduce Kepler data on quasars, as it differs significantly from the standard analysis used for the primary Kepler science goal of exoplanet transit detection.   We cover the post-processing analysis, including Palomar Observatory imaging and photometry, in \S 5.  In \S 6 we discuss our results, and finally summarize our conclusions in \S 7.

\section {Variability Signatures of  Disks and Jets}
\subsection{Emission from Accretion Disks}
Fluctuations emerging directly from the surfaces of ADs or from coronae above them can produce rest frame variations no faster than about $t_{var} \simeq GM_{BH}/c^3$ or $\simeq 8$ minutes for  $M_{BH} = 10^8 M_{\odot}$.  
The fastest real variations can provide a  way of measuring the lower limits on masses of the SMBHs. Most X-ray and optical variability of the best studied cases, Seyfert galaxies and radio-quiet quasars, is ``red noise", i.e., where the PSD arising from the Fourier transform of the light curve is characterized by $P(f) \propto f^{\alpha}$, with $\alpha < 0$, below some break frequency, $f_b$; above $f_b$ the PSD is usually dominated by measurement errors with a white noise character ($\alpha = 0$) or perhaps by white noise with an astrophysical origin.  The PSD can also yield the size of the largest emitting region if there is a flatter slope, often $\alpha \sim -1$ at low frequencies and $\sim -2$ at intermediate frequencies \cite[e.g.,][]{markowitz03} before perhaps turning to $\sim 0$ at the highest frequencies.  Variations yielding red-noise can be produced if energy is released over a wide range of time-scales characteristic of different orbital periods and turbulent transport in ADs \cite[e.g.,][]{mw93}.

\subsection{Previous indications of quasi-periodic variations}

Recently, detections of quasi-periodic oscillations (QPOs) in five AGN have been made, mostly in X-rays.  The best case is that of the Narrow Line Seyfert 1 galaxy RE J1034$+$396 \citep{Gierlinski08},  where a period of a little more than 1 hour was detected. The presence of optical QPOs has been suggested for the high declination BL Lac object, S5 0716$+$714, for some time \citep[e.g.,][]{quirrenbach91}.  A wavelet analysis of archival optical spectra of S5 0716$+$714 indicated the presence of five nights with QPOs  present at $\geq 0.99$ probability, with central periods ranging between $\simeq 25$ and $\simeq 73$ minutes \citep{gsw09}.   New observations of this same blazar led to an even stronger indication of an $\sim$15 min QPO \citep{rani10}.  

The simplest QPO model assumes that a single hot-spot  
in the inner portion of the AD is responsible, and that radiation from this disk is directly detected, e.g., when an FSRQ is in a faint, quiescent state.  Given a measured period, $P$ (in seconds) one can estimate the SMBH mass $M$ via

\begin{equation}
\frac {M} {M_{\odot}} = {\frac {3.23 \times 10^4 ~ P} {(r^{3/2} + a)(1+z)}},
\end{equation}
where $r$ is the hot-spot distance in units of $GM/c^2$, $a$ is the BH spin-parameter and $z$ is the redshift \citep{gsw09}.
The shortest periods are obtained for the innermost stable circular orbit, and the strongest radiation emerges near there.  For low redshift objects, a 5 hour period, which would be very difficult to detect from ground-based observations, but rather easy for Kepler to find, corresponds to SMBH masses of $4.0 \times 10^7$ and $2.5 \times 10^8$ M$_{\odot}$, for Schwarzschild and extreme Kerr BHs, respectively.  
In practical terms, we need to observe for many months to be sensitive to ADs surrounding billion-solar-mass black holes because any  large hot spots in ADs are probably both rare and short-lived,  lasting no more than dozens of orbits ($\sim$ 2 weeks).   

\subsection{Emission from Jets}
For core-dominated flat spectrum radio quasars,  where the emission is almost certainly dominated by jets  within several degrees to our line of sight, the radio--X-ray variability can result from changes in a variety of physical parameters: the synchrotron-emitting population of particles; the conditions under which they move; the orientation to line of sight; or the Doppler 
factor, $\delta \equiv [\gamma (1-\beta \cos \theta)]^{-1}$, with $\beta = V/c$ the velocity of the shock through the jet, and $\theta$ the angle between the observer's line of sight and the jet axis.. Major flares arise from new shocks passing through the jets \citep{mg85}, but smaller fluctuations could arise from turbulence behind those shocks \citep{marscher08} or ``mini-jets" \citep{giannos09,giannios10,nalewajko11}. Alternatively, variations in the magnetic fields, changes in the ambient medium, and changes in the injected particle distribution can also cause variations in emitted light. 
The light travel time across the radiating region, modified by the Doppler factor, is the characteristic time scale.
If these variable emissions arise from shocks in relativistic jets, as is expected to be the case for FSRQs in high states, then the observed $t_{var} \simeq   \Delta R (\delta c)^{-1}$ \citep{gk03}. Here $\Delta R$ is the physical size of the emitting region in its rest frame (roughly a jet diameter) and $\delta$ the Doppler factor.    In this case, variability timescales can constrain the jet's  $\delta$  \citep{jorstad05}.    
Constraints on the jet velocity and viewing angle can be particularly tight if other data, such as apparent superluminal motions of radio knots, where $V_{app} = V \sin \theta (1-\beta \cos \theta)^{-1}$, can be detected through VLBI.  Our 4 FSRQ targets are all VLBA calibrators with compact radio structures (e.g., [HB89] 1924+507=4C50.47, a member of the CJF survey in \cite{britzen2008}).

A leading model for any jet-based quasi-periodic variations is  the intersection of an outward propagating shock with a helical structure within the jet \citep{rani09}.   Even small deviations in the viewing direction result in significant changes in flux.   The period of a QPO and its stability between events in a given source is a mode discriminator.  Helical structures in a jet are relatively long-lived, so that QPOs associated with them should have similar periods during successive events.   But if QPOs involve turbulent cells behind the shock in the jet \citep{marscher08,rani09}, short-lived dominant eddies are expected to give rise to different periods.

\subsection{Duty cycles and timescales of variable emission from  relativistic jets}

Dramatic variability is a defining characteristic of blazars; however, the duty cycles of activity are still not fully characterized, since the number of objects subject to sustained monitoring from the ground is not large.  FSRQs vary on timescales of minutes to decades, with the most variable (e.g., 3C\,279) ranging two magnitudes from their average level and up to five magnitudes above their low quiescent level \citep{Kartaltepe07}.  
Microvariability, at the level of 0.03 mag over the course of a night (typically 6 hours) is quite common, with duty cycles around 50\% for blazars \citep[and references therein]{gk11,carini07,goyal12}.
When the smallest variations detectable from the ground, at about 0.01 mag per night, are included, the blazar duty cycle rises to $\sim$ 80\%.  The duty cycle for similar small variations from normal radio loud (steep spectrum) quasars, for which jets are present, but not strongly Doppler boosted, is closer to 20\% \citep[e.g.,][]{sagar04,stalin04,ramirez09}, which might be expected.
More surprisingly,  radio-quiet QSOs  show a similar duty cycle to radio-loud quasars, indicative of variability arising from accretion disks or perhaps from jets that are present only on nuclear scales   \citep[e.g.,][]{gk03,ramirez09}. For one example, CTA\,102 is a well studied, but otherwise typical, highly polarized bright ($V \sim 15$) blazar.   
\cite{osterman09}  found variations as large as 0.1 mag in 15 minutes, with typical slopes of 0.01 mag/hr. 

Structure function analysis is used to characterize the variability timescale and infer the physical scales when the Doppler factor and orientation to our line of sight are also measured.  \cite{bachev12} monitored the short term optical variability of 13 highly variable blazars (not a complete sample) over the course of five years. They found surprisingly low variability: several objects varied slowly and smoothly at rates of up to $\sim 0.1$ mag per hour, but many displayed no short term variability. They quote only a $\sim 2 \%$ chance of observing variability of more than 0.1 mag~per hour during their observations.

\section{Sample Selection}

We selected objects with strong compact cores by searching the NRAO VLBA Calibrator List\footnote{http://www.vlba.nrao.edu/astro/calib/} for radio sources brighter than 100 mJy on $\sim$400 km baselines at 2.3 GHz in the Kepler field of view. We found four AGN that met these criteria.  The properties of these targets are shown in Table \ref{target-list}; for convenience, we refer to them throughout this paper as Objects A-D. 
Three objects are classified as quasars and have radio spectral indices $\alpha$ between 0.0 and 0.32 (where $S \propto \nu^{-\alpha}$); the fourth object is a tentative Seyfert 1.5 galaxy which has $\alpha = 0.75$ , typical of steep spectrum (extended) radio emission.  We also observed the archetypal powerful radio galaxy, Cygnus A, with various custom apertures coordinated by Michael Fanelli at the Kepler Science Operations Center.   Analysis of Cygnus A requires special treatment due to its extended optical structure and crowded foreground stars; we expect to present data and results in a subsequent paper. We observed all five targets in Kepler Guest Observer Cycles 2 and 3, corresponding to Quarters 6-13 (two years of almost continuous monitoring in 2010-2012, Kepler GO Programs GO20018 and GO30010), and we report these results and our analysis here.  Further data on our targets (Cycle 4) is currently being acquired. 
 
\section{Data Reduction and Analysis}

\subsection{Overview }  The Kepler focal plane has 21 modules each with two 2200$\times$1024 pixel CCDs for astronomical observing and 4 CCDs for fine guidance. The images are out of focus to improve sensitivity to periodic signals for extrasolar planet detection around bright stars. Thus the signal from even a point source is spread over several 2.98'' pixels. The signals from pixels in  an aperture whose size is a function of anticipated magnitude are sent down by the Kepler spacecraft \citep{bryson2010}.  For faint targets such as our AGN (V magnitudes 17.8-18.6), the signal is several hundred electrons per second, a far different regime for instrumental response (and artifacts) than for the bright stars  ($V < 10$) searched for extrasolar planets.  We observed all of our targets in Long Cadence (LC) 30-minute mode, yielding $\sim$4500 samples each quarter per target. In addition, we observed our targets in Short Cadence (SC) 1-minute mode for  a single quarter each year ($\sim$~125,000 samples).   Object B falls on dead CCD module \#3 for one quarter of each year (Q8 and Q12 during Cycles 2 and 3), so we have only 6 quarters of data for it.  
In LC mode for Object D at $V=18.4$ in the Kepler Input Catalog, the measured per-cadence noise level ($1 \sigma$)  was  0.9\%.  In SC mode, the per-cadence noise level at $V=18.4$ was 3.4\%. A journal of the observations is given in Table 2.  

The standard Kepler pipeline data have the day-to-week-scale drifts removed because the pipeline is optimized for extrasolar planet detection, which also removes real astrophysical brightness variations of the same timescales. Therefore, we used pre-pipeline data, ``SAP FLUX" , in our analysis. This was also the approach taken by \cite{mushotzky11} and \cite{cr12} in their analyses of Kepler data for brighter AGN.    This choice  means that some of the long term variations that are seen in the data are instrumental and not intrinsic to the source, but we can also be confident that we are not discarding any of the short term data trends that are intrinsic to the source.   We removed the largest contribution of this effect on the PSDs we computed by performing end-matching (discussed below) on each quarter (for LC data) or each month (for SC data). 

\subsection{Instrumental Effects: Differential Velocity Aberration, Aperture Optimization, and Nonlinear Amplifier Electronics}

We note that there is no precise way to translate the observed signal in electrons/second to optical magnitude because this was not a mission design feature; Kepler was designed to measure very precise changes in relative brightness, not very precise absolute brightness. The light curves formed by Kepler long term data are affected at the few percent level by {\it differential velocity aberration} which originates in the motion of the spacecraft in its annual orbit around the sun. As the spacecraft moves, it keeps a constant field of view in the constellation Cygnus, and the angle formed by its velocity is constantly changing. 
This causes a target's PSF to move with respect to the detector array, and because a limited number of optimal aperture (``postage stamp") pixels is downloaded, each quarter's light curve would have a characteristic rising or falling slope or a concave or convex shape, superposed on other effects such as thermal drift.   The finite number of pixels used for photometry is also responsible to the steps seen between quarters; after a spacecraft roll, an object falls on a different CCD and in a different location with respect to the pixel array.  
Five of the 16 CCDs that our four AGN land on each year are affected by an unstable amplifier electronics problem which causes a quasi-periodic signal to move across the CCDs (see \citealt{Kolodziejczak10} for technical details).   This Moir\'e effect, so-named because of its resemblance to a Moir\'e pattern, affects faint sources more than bright sources because it is additive, not multiplicative.  As of writing, no models had been developed to remove the effect from faint source signals.  Affected CCDs and modules are listed in Table 13 of the Kepler Instrument Handbook; see also their Figure 24 \citep{kepler2009a}, and for module number identification, see Figure 2-1 of the Kepler Archive Manual, \citep{kepler2009b}.    These ``Moir\'e patterns'' look like ripples with timescales of days, and are visible in data from some quarters, as listed in Table 2.  The vast majority (80\%) of our data are of excellent quality and are {\it  not} affected by this problem. 

\subsection{Light Curves}

We show light curves for all four AGN in Figure 1.   The time axis is given in Barycentric Julian Date (BJD), the standard Kepler time unit which is referenced to the Solar System barycenter; it differs from the more familiar (heliocentric) Julian Date by $ \pm 4$ seconds (the recently discovered absolute timing error in the FITS headers (M. Still, Kepler Blog of 12 Dec 2012\footnote{http://keplergo.arc.nasa.gov/Blog.shtml}) has no effect on our studies). The repeated annual patterns of differential velocity aberration are the clearest features, producing long term smooth variations.  

\section{Light Curve Analysis and Results}

\subsection{Flares in CGRaBS J1918+4937 and MG4 J192325+4754}

We found a $\sim 7$\% rise in the flux of CGRaBS J1918+4937 (object C) during Q10  which lasted about five or six days  around BJD=2455755 (Fig.~2). We confirmed that there was no plausible contamination source, e.g., a nearby star that might have flared and been incorporated into the pixels included in its photometry. This variation shares no characteristics with known artifacts, hence, we conclude that it is real.  Object C also showed three consecutive fluctuations of $\sim 3\%$ amplitude during a span of about 25 days during Q9.  Other variations in Object C during Q7 and Q11 might be real, but because there was some Moire effect present in those quarters, we cannot make that attribution.   Object A (MG4~J192325+4754) showed several  significant flares of $\sim 4-6$\% size over a period of 10 days in Q12.  This quarter is one of the two for that AGN that should not have been affected at all by Moir\'e problems but this quarter was affected by solar coronal mass ejections \citep{kepler2012}; however, the interval in question was well away from those disruptions and also well away from the monthly gaps and glitches.  Neither object B nor D showed any short-term variability of any type during any quarter.  

\subsection {Search for Brief Flares}

No fast astrophysical flares (shorter than a few hours) were detected in any of our targets. Our search method was as follows. Two of us  independently searched the SC and LC light curve data for flares.  We used the {\it Pyke} task {\it Keppixseries} \citep{still2012} to examine the data in each target's pixel; then we used {\it topcat} \citep{taylor2005} to find the flag values for anomalously high data points, identified in the Kepler Archive Handbook, Table 2-3.  Both the two significant  (defined at 4 to 6 consecutive high points each) flares turned out to be instrumental and not astrophysical. The problems were flagged in the FITS files as ``coarse pointing" but were actually probably ``argabrightening" \citep{jenkins2010} that was not recognized as such by the pipeline software.  During Q12, strong solar coronal mass ejections caused spacecraft coarse pointing problems that were manifested as many anomalously high data points along with several anomalously low readings \citep{kepler2012}.

\subsection{Power Spectral Densities}

We constructed PSDs for all the Long Cadence data, then fitted slopes to the low frequency and high frequency portions of those PSDs.    
We  computed PSDs  with a Discrete Fourier Transform using {\it Period04} \citep{lenz05}, both for the original SAP count rates and our ``corrected" data, which underwent the following processing steps that fixed bad data points and would also allow us to use a fast Fourier transform on the light curves (see Fig.~2).   First, monthly ``glitches'' were removed.  These instrumental effects from the temporary changes in CCD sensitivities produced by the temperature changes resulting from reorientation of the spacecraft required to send back data each month often were at the level of 2-5\% of the flux.  We modeled them as exponential decays, with amplitudes and time constants as free parameters and then subtracted (or added) the corrections needed to produced smooth post-glitch curves.  The typical decay time was 0.8 to 1.5 days.  
Second, the gaps produced by the monthly data downloads, along with the occasional other gaps in the data (usually just a few dozen isolated points per quarter, occasionally two to five contiguous points) were filled in by fitting third order polynomials to the data on either side of the gaps and then adding noise derived from the local standard deviation.  In the case of Q12 data, which was affected by coronal mass ejections, these additional gaps lasted up to 4 days.  Third, single point outliers more than $4 \sigma$ from the local mean were removed and replaced with points generated from that local mean with a random dispersion based on that $\sigma$ value.  Finally, in order to minimize errors in the PSD produced by the low-frequency instrumental drifts discussed above and to take into account the fact that we do not have an infinitely long data train, we ``end-matched'' each light curve \citep[e.g.,][]{fou85,mushotzky11}, removing a linear trend so that the last several points are at the same level as the first several points in each quarter.  

The PSD slope is determined from equally weighted points in the PSD. The PSD is not uniformly populated with data; there is far more data at the high frequency end  than the low frequency end.   The slopes at the high frequency end are invariably consistent with 0, or white-noise, and our fits were done so as to force the high frequency PSD to a zero slope.  Then the low-end PSD slope was found via a least squares procedure computed starting at the lowest frequency points until that slope started to flatten dramatically, which was indicative of the location of the break frequency.  
The results are summarized in Table 3, which includes results for the original SAP data, end-matched SAP, corrected SAP and corrected-end-matched SAP.  The slopes based on original SAP counts are usually between $-1.5$ and $-2.0$, but vary somewhat from quarter to quarter for the same source and from source to source, with Object D usually having a shallower slope than the other three.   
The corrected data yielded very similar slopes to those of the original data, though they were usually slightly steeper.
 It is very difficult to estimate the actual errors in the PSDs \citep[see][]{vaughan2003}, but the quarter-to-quarter differences between slopes of the same object are usually characterized by standard deviations of 0.1-0.2, so the uncertainties involved in the computations should be no larger than this, since there is probably some degree of intrinsic variation between quarters.

The slopes found using end-matched SAP data are almost always shallower than those using the pure SAP counts, which is expected since some of the long-term drift, which yields power at the lowest frequencies, is removed with end-matching.    They are usually between $-1.2$ and $-1.9$. Note that if the intrinsic PSDs were actually steeper than $-2.0$, ignoring end-matching would tend to drive the estimated slopes toward around $-2.0$, while incorporating it provides much closer estimates of the actual slopes fed into simulations \citep{fou85}.  The fact that end-matching typically slightly flattens the slopes indicates that the intrinsic slopes were not steeper than $-2.0$.    We show typical examples in Figure 3.  

We also fitted the Short Cadence data and give, in Table 4, the results for  the original SAP data, the corrected SAP data and the corrected, end-matched SAP data; however, the last two values were always the same to at least two significant figures.  For each quarter for each source, the break frequencies in the SC data turned out to be essentially the same as in the LC data,  which reinforces our confidence in the fitting.

\subsection {Palomar Observatory Photometry} 

Blazars can be ``bluer when brighter" or ``redder when brighter".  Accordingly, we obtained optical and near-infrared photometry from CCD observations of our targets at the Palomar Observatory 60 inch telescope in September, November and December 2010, 
during Kepler Q7.
We used Sloan $g' r' i' z' $  filters to evaluate the overall spectral shape within the very broad Kepler bandpass (Kepler's CCDs, which have no filters, are sensitive to visual wavelengths between from $4200 - 9000$ \AA).  Data on the target AGNs and five to six nearby comparison stars in each field were reduced with the aperture photometry tasks in {\it IRAF}.  Data were calibrated using the stellar magnitudes in the Kepler Input Catalog \citep{brown11}. Dereddening was applied using values from the NASA Extragalactic Database (NED).  Results are given in  Table 5.  Objects A and B have nearly power-law spectra consistent with nonthermal emission, with  $\alpha = 1.5$ and  $2.9$, respectively  for S $\propto \nu ^{-\alpha}$). For comparison, blazar 3C\,454.3, observed on 23 Nov 2010 in the course of another Palomar program, had a spectral index $\alpha = 2.0$. In contrast, Object C has a nearly flat spectrum (bluer than the other three objects)  consistent with a blue continuum spectrum showing a single emission line,  MgII at 2798\AA,  (\citealt{healey2008}; FITS spectrum provided by M. S. Shaw, personal communication). In Object C, therefore, the optical emission could have a significant contribution from  thermal Big Blue Bump emission, but disentangling the relative contributions of nonthermal and thermal emission requires high quality spectrophotometry which is beyond the scope of this paper. Object D has a spectrum peaked at $i'$ band, but the large uncertainty at $z'$ band is sufficient to encompass a power law similar to those seen in Objects A and B.
   
\subsubsection{Upper Limits on Variability}

None of the four AGN varied significantly during the Palomar observations in September through December 2010. 
At $r'$ band, the brightness of the four targets was similar to those observed at the 1.2-m telescope at Mt. Hopkins in 2003-2008, as listed in the Kepler Input Catalog \citep{brown11}.   Differences of  $\sim 0.1-0.4$ magnitudes in three targets at $g', i'$ and $z'$ were observed between the Kepler Input Catalog (KIC) measurements  and the Palomar measurements; the reddest AGN (whose spectrum is most similar to the red stars used for calibration) showed no discrepancy. Three quasars are always fainter in blue bands and brighter in red bands than they were in the KIC. We attribute the Palomar-KIC differences to the effects of reddening and CCD-filter responses to the dissimilar spectra of quasars vs. stars. Observations at $z'$, where the discrepancy is the largest, are most affected by site-variable atmospheric water vapor \citep{fukugita1996}, moreover, different CCDs have different $z'$ responses due to sensitivity variations at the longest wavelengths. 

\subsubsection{Morphology}

Three of the four targets were unresolved in all four Sloan bands. Object B, the lowest redshift target,  was resolved by $g'$-band imaging into a central point source embedded in diffuse emission typical of host galaxies. With ground based seeing of $1.^{\prime \prime}5$ and with the target at redshift 0.513, no structural details were expected or revealed; the fuzzy host galaxy emission in the $g'$-band image was estimated at approximately 6 arcsec extent using {\it IRAF} tasks.   In the other Sloan bands, Object B was much more strongly dominated by the central nuclear source with comparatively weak diffuse emission detected in radial profile plots. It was tentatively classed as a Seyfert 1.5 galaxy based on observations of a spectrum with a single emission line, H$\alpha$, by \citet{hen97}. 

\section{Discussion}

\subsection{Flares and the Overall Shape of the Light Curves}  
In 30 quarters of data, the strongest flare we saw was an excursion  of $\sim 7\%$ amplitude in Object C during Q10, and lasting about 5 days.     Object A also several  significant flares of $\sim 4-6$\% size over a period of 10 days in Q12. Object C showed three excursions of $\sim 3\%$  in Q9. Objects B and D showed no days-scale variability  during any quarter.   The flares could be caused by magnetic reconnection events or random fluctuations within the jets or by brightening in the accretion disk. No microvariability on timescales of minutes to hours was detected, possibly because the short-cadence noise levels of 2-3.4\% were too high; ground-based observations show levels of 0.01-0.03 magnitudes ($\sim 1-3\%$) are common in blazars, but rare in other classes of AGN, as reviewed in our Introduction.   

We did not detect any astrophysical QPOs, which was unsurprising given the rarity of such events.
Given the very long, consistent, light curves produced by Kepler, we greatly improved the chance of finding QPOs, if they were to be present on timescales of tens of minutes to weeks and sufficiently strong.   We quantified our sensitivity to sine waves injected at  a range of frequencies and amplitudes: in a blind study where one of us added such signals to real data another of us discovered that he could detect them in the PSDs at a level of $\sim$ 3\% of the original amplitude for 18th magnitude targets, but could not do so for periodic signals at 1\% amplitude.

The comparative smoothness of the Kepler light curves within each quarter is consistent with the finding by  \cite{bauer2009} that only 35\% of blazars showed $V > 0.4 $ magnitudes of variation over 3.5 years, as observed with intermittent sampling during the Palomar Quest survey. 
Longer term optical variability of quasars has traditionally involved frequent monitoring of a modest number of objects from individual observatories; those observations indicate that essentially all quasars do vary substantially over decades, and changes over a year or two are expected \citep[e.g.,][]{pica88,hawk02}.    Averages of such monitoring can yield statistical measures of quasar variability through the use of structure functions \citep[SFs, e.g.][]{hughes92,coll01}.  Such monitoring, however, can be biased in favor of ``interesting'' objects, with those known to be variable being covered more frequently and some ``uninteresting'' objects can be dropped from the programs in favor of newly discovered variable objects.  With the advent of massive sky surveys that revisit patches of the sky, such as the Sloan Digital Sky Survey \citep[SDSS,][]{york00}, a new approach has become possible: one can compute average variability estimates for tens of thousands of quasars divided into substantial bins by their redshifts, luminosities and other properties \citep[e.g.][]{vanden04,devries05}.  Recently, structure function (SF) analyses of quasar variability using SDSS and the Palomar Sky Survey have led to estimates of quasar variability and a damped random walk model that seems to explain its main properties \citep{macleod10,macleod12,ruan2012}.  One conclusion of that work particularly relevant to our study is that radio-loud quasars are slightly more variable than radio-quiet ones.    Specifically, the average $SF_{\inf}$,  a measure of the strength of variability,   was around 0.26 mag (in $g'$) for the whole sample while the average for the radio-loud sample was 0.34 mag \citep{macleod10}, where the redshifts in both categories were quite similar.  Our densely sampled Kepler data on four AGN complement those programs with baselines of years that looked at very large numbers of objects much less frequently.
        
\subsection{Power Spectral Densities} 

The results of our power spectral density (PSD) analyses are shown in Table 3 (long-cadence data) and Table 4 (short-cadence data).  As discussed in \S 5.3, possible systematic errors in slopes are minimized by the corrections for (post-monthly-download) thermal drift  and end-matching (last column of Table 3). The power-spectra for our quasars are dominated by red noise at lower frequencies and white noise at higher frequencies.   We interpret the bulk of the red noise as intrinsic to the quasars;  the white noise is instrumental. 
We find mean slopes between $-1.2$ and $-1.8$ for the red-noise portion of the PSDs for our four AGN.     
These slopes varied from quasar to quasar and, to a somewhat lesser extent, from quarter to quarter for the same object.   Object D displayed  the flattest slopes on the whole, with those for Object B were somewhat steeper but distinctly  shallower than Objects A and C.   Object B, tentatively classed as a Seyfert 1.5 galaxy, is the object in which our Palomar imaging detected the faint host galaxy surrounding the bright central point source, hence, the Kepler-detected optical emission had a small extra contribution from the host galaxy which could damp down any variation and flatten the resulting  PSD.  Among our four objects, there was no trend for PSD slopes to vary with redshift.  The observed PSD slopes are consistent with those of turbulence in shocks in a relativistic jet or in an accretion disk \citep{mt91,mw93}. Models of relativistic turbulence which may apply to our quasars are being developed in the context of gamma ray bursts \citep{zrake2013} and references therein.

Kepler measurements of the brighter Seyfert galaxies analyzed by \citet{mushotzky11} have substantially steeper slopes (between $-2.6$ and $-3.3$) for the PSDs of their 4 Seyferts, and they also employed end-matching.   Their objects were all substantially brighter than our targets (by factors of about 1.5 to 15) and so they could easily detect smaller fractional variations of 0.1--1\%.  Given that our objects are at much greater distances, and therefore substantially more powerful, despite their lower fluxes, there might be some physical reason for the discrepancy.  Their results of very steep PSD slopes were surprising, because long-term X-ray measurements of other Seyferts yielded PSD slopes hat are almost always  shallower than $-2.0$ \citep[e.g.][]{ede99,markowitz03,mar09,lac09}.   Further, for the brightest of one of these Seyferts,  II ZW 229.015, for which \citet{mushotzky11} found slopes between $-2.96$ and $-3.31$, a reanalysis by a different method, coupled with ground-based observations   yielded a best-fit shallower slope around $-2.83$ \citep{cr12}.   Our independent analysis of data for that object following the correction and end-matching procedures used in this paper produced slopes consistent with $-2.2$.  We note that \citet{mushotzky11} remark that ``PSD analyses are notoriously susceptible to analytical systematics (see, e.g., \cite{vaughan2003})". Users of PSDs are on firmer ground when comparing ``apples to apples", that is, PSDs for different objects are internally consistent when computed by each group's analysis techniques, but PSDs generated by different groups' analysis techniques are much more difficult to compare.

The break frequencies at which the red noise and white noise contributions are equal  were between $\sim 10^{-4.2}$Hz and $10^{-4.7}$Hz, corresponding to 0.2 to 0.6 days.     Increasing the instrumental noise raises the white noise ``floor" in the PSD diagrams, shifting the intersection of red noise and white noise to longer timescales (to the left in our figures).  Kepler measurements of the brighter Seyfert galaxies analyzed by \citet{mushotzky11}, which have smaller fractional noise, also have break frequencies of about 0.25 days.

\subsection{Future Work} We are currently acquiring additional long cadence data on each target. In future work, we will apply principal components analysis and cotrending basis vectors to the long cadence data to separate out  common instrumental long term trends (such as differential velocity aberration and thermal drift, not Moir\'e effects) in the data. These techniques may allow us to patch together individual quarters, forming light curves two or more years long for three targets and several three-quarters light curves for the remaining target that falls on a dead CCD during one quarter each year.    Our goal is to detect lower break frequencies in the PSDs to find the largest physical scale as was done for II Zw 229.015 by \cite {cr12} who used ground-based data to bridge quarterly amplitude jumps in their Kepler light curves.

\section{Summary}

We have observed three flat spectrum radio quasars and one radio-loud AGN (tentatively classed as a Seyfert 1.5 galaxy) for two years with Kepler.   We find power spectral densities in the light curves that vary from source to source and quarter to quarter for the same source, ranging from $\alpha = -1.2$ to $-1.8$.  We find several clear, isolated flares over a few percent amplitude which are of astrophysical origin.  We do not detect any of the expectedly rare quasi-periodic oscillations as could be generated by accretion disks or helical jet features. The power spectral densities measured agree with models for the observed variability originating in turbulence behind a shock in the jets or in the accretion disks.   These observations provide beautiful sets of high quality data.  Future work on combining data across quarterly boundaries and extending the observations  for the life of the Kepler mission could provide us with estimates of the largest relevant scales of the physical processes producing the variability.  With the further development of shocked jet models including turbulence (e.g., Marscher, in preparation), these types of continuous light curves may distinguish between jet turbulence and accretion disk fluctuations as the dominant source of quasar optical variability.

\section{Acknowledgements}

We thank Martin Still and Tom Barclay (Kepler Science Center, NASA Ames Research Center), Jeff Kolodziejczak (Kepler Mission, Marshall Space Flight Center), Rick Edelson (University of Maryland Baltimore County), Mike Carini (Western Kentucky University) and Paul Smith (Steward Observatory) for helpful discussions.  We acknowledge support from the NASA Kepler Guest Observer Program through grants to A. E. Wehrle  (NNX11B90G and NNX12AC83G). The TCNJ group acknowledges additional support from the Mentored Undergraduate Summer Experience program. This work made use of {\it PyKe},  a software package for the reduction and analysis of Kepler data. This open source software project is developed and distributed by the NASA Kepler Guest Observer Office. This research has made use of the NASA/IPAC Extragalactic Database (NED) which is operated by the Jet Propulsion Laboratory, California Institute of Technology, under contract with the National Aeronautics and Space Administration.  Some of the data presented in this paper were obtained from the Mikulski Archive for Space Telescopes (MAST). STScI is operated by the Association of Universities for Research in Astronomy, Inc., under NASA contract NAS5-26555. Support for MAST for non-HST data is provided by the NASA Office of Space Science via grant NNX09AF08G and by other grants and contracts.

{}
% --------------------------------------------------------------------------------------------------------
% --------------------------------------------------------------------------------------------------------
\begin{deluxetable}{cccccccc}

\tabletypesize{\scriptsize}

\tablewidth{175mm}

\tablecaption{Kepler AGN Monitoring Target List\label{target-list}}

\tablenum{1}

\tablehead{

\colhead{Object} & \colhead{Name} & \colhead{Kepler Input} & \colhead{Right} & \colhead{Declination} & \colhead{Kepler Input} & \colhead{Redshift} & \colhead{Radio} \\ 

\colhead{Designation} &  & \colhead{Catalog} & \colhead{Ascension} &  & \colhead{Catalog} & & \colhead{Spectral} \\ 

&  & \colhead{Number} &  &  & \colhead{Magnitude} &  & \colhead{Index\tablenotemark{a}} \\ 

\colhead{} & \colhead{} & \colhead{} & \colhead{(hh:mm:ss.s)} & \colhead{(dd:mm:ss)} & \colhead{} & \colhead{} & \colhead{} } 

\startdata
A & MG4 J192325+4754 & 10663134 & 19:23:27.24 & 47:54:17.0 & 18.6 & 1.520 & 0.32 \\
B & MG4 J190945+4833 & 11021406 & 19:09:46.51 & 48:34:31.9 & 18.0 & 0.513 & 0.75 \\
C & CGRaBS J1918+4937\tablenotemark{b} & 11606854 & 19:18:45.62  &  49:37:55.1 & 17.8 & 0.926 & 0.00 \\
D & [HB89] 1924+507 & 12208602 & 19:26:06.31  &  50:52:57.1 & 18.4 & 1.098 & 0.19 \\
\enddata

\tablenotetext{a}{Radio spectral index obtained from VLBA Calibrator website, defined between 2.3 and 8.3 GHz or 2.3 and 8.6 GHz with S $\propto \nu^{-\alpha}$}

\tablenotetext{b}{Kepler Input Catalog incorrectly indicates that this target is a star with contamination 0.73, but we have verified it is an isolated quasar.}

\end{deluxetable}

% ----------------------------------------------------------------------

\begin{deluxetable}{ccc}

\tabletypesize{\scriptsize}

\tablewidth{150mm}

\tablecaption{Journal of Kepler Observations of AGN\label{journal}}

\tablenum{2}
\tablehead{

\colhead{Object} & \colhead{Long Cadence\tablenotemark{a}} & \colhead{Short Cadences\tablenotemark{c}}  \\ 

& \colhead{(Kepler quarters)\tablenotemark{b}} & \colhead{(Kepler months) }  

}

\startdata
A & Q6*, 7**, 8, 9**, 10*, 11**, 12, 13** & Q6.1, 6.2, 6.3, 11.1, 11.2, 11.3 \\
B & Q6, 7, 9, 10, 11, 13 & Q6.1, 6.2, 6.3, 11.1, 11.2, 11.3 \\
C & Q6, 7*, 8, 9, 10, 11*, 12, 13 & Q6.1, 6.2, 6.3, 11.1, 11.2, 11.3  \\
D & Q6*, 7, 8, 9, 10*, 11, 12, 13 & Q6.1, 6.2, 6.3, 10.1, 10.2, 10.3  \\
\enddata

\tablenotetext{a}{Targets are observed on a given Kepler detector for one quarter, using 4 detectors during a year. Start-stop dates for quarters are given at: http://keplergo.arc.nasa.gov/ArchiveSchedule.shtml.  For our Q6-13 data, each quarter returned approx.\ 4100-4700 30-minute cadences. Object B falls on dead CCD \#3 during Q8, Q12, etc.}

\tablenotetext{b}{Some data are affected by a Moire fringe-like time-dependent instrumental fluctuation in amplitude that is very hard to calibrate. * denotes likely fringing; ** denotes fringing obvious by inspection of the raw data.}

\tablenotetext{c}{For our Q6, 10 and 11 data, each month returned approx.\ 40,000 1-minute cadences.}
\end{deluxetable}
% ----------------------------------------------------------------------

\begin{deluxetable}{cccccccc}

\tabletypesize{\scriptsize}

\tablecaption{AGN variability power spectral densities (PSDs) measured with long-cadence Kepler data}

\tablenum{3}

\tablehead{

& & & & \multicolumn{2}{c}{SAP (raw) data} & \multicolumn{2}{c}{Corrected SAP data} \\

\colhead{Object} & \colhead{Kepler} &  \colhead{Quarter} & \colhead{Moir\'e} & \colhead{Slope} & \colhead{Slope} & \colhead{Slope} & \colhead{Slope} \\

\colhead{Designation} & \colhead{ID} &  &  \colhead{level} & \colhead{(original)} & \colhead{(end-matched)} & \colhead{(original)}  & \colhead{(end-matched)}
} 

%% All data must appear between the \startdata and \enddata commands
\startdata
A & 10663134 & 6 & Medium & -1.9 & -1.7 & -2.0 & -1.6 \\
A & 10663134 & 7 & High & -2.1 & -2.1 & -2.3 & -1.3 \\
A & 10663134 & 8 & None reported & -2.0 & -1.9 & -2.1 & -2.0 \\
A & 10663134 & 9 & High & -1.7 & -1.3 & -1.6 & -1.4 \\
A & 10663134 & 10 & Medium & -1.5 & -1.6 & -1.6 & -1.4 \\
A & 10663134 & 11 & High & -2.0 & -1.7 & -2.1 & -2.0 \\
A & 10663134 & 12 & None reported & -1.8 & -1.9 & -2.0 & -1.8 \\
A & 10663134 & 13 & High & -1.9 & -1.7 & -1.8 & -1.8 \\
\tableline
Mean\tablenotemark{a} &  &  &  & -1.8 & -1.8 & -1.9 & -1.7 \\
SD &  &  &  & 0.2 & 0.2 & 0.2 & 0.3 \\
\tableline
B & 11021406 & 6 & None reported & -1.9 & -1.6 & -2.0 & -1.4 \\
B & 11021406 & 7 & None reported & -1.8 & -1.3 & -1.9 & -1.2 \\
B & 11021406 & 8 & None reported & \nodata & \nodata & \nodata & \nodata \\
B & 11021406 & 9 & None reported & -1.5 & -1.7 & -1.8 & -1.8 \\
B & 11021406 & 10 & None reported & -1.8 & -1.0 & -1.5 & -0.9 \\
B & 11021406 & 11 & None reported & -1.6 & -1.7 & -1.8 & -1.3 \\
B & 11021406 & 12 & None reported & \nodata & \nodata & \nodata & \nodata \\
B & 11021406 & 13 & None reported & -1.8 & -1.6 & -1.9 & -1.7 \\
\tableline
Mean &  &  &  & -1.7 & -1.5 & -1.8 & -1.4 \\
SD &  &  &  & 0.2 & 0.3 & 0.2 & 0.3 \\
\tableline
C & 11606854 & 6 & None reported & -1.9 & -1.4 & -2.0 & -1.6 \\
C & 11606854 & 7 & Medium & -1.5 & -1.9 & -1.6 & -2.0 \\
C & 11606854 & 8 & None reported & -1.8 & -1.2 & -2.0 & -1.6 \\
C & 11606854 & 9 & None reported & -1.9 & -1.9 & -1.9 & -2.0 \\
C & 11606854 & 10 & None reported & -1.9 & -2.4 & -2.1 & -2.1 \\
C & 11606854 & 11 & Medium & -1.8 & -2.0 & -2.0 & -2.0 \\
C & 11606854 & 12 & None reported & -1.8 & -1.7 & -1.9 & -1.7 \\
C & 11606854 & 13 & None reported & -1.8 & -2.0 & -2.0 & -2.0 \\
\tableline
Mean &  &  &  & -1.9 & -1.8 & -2.0 & -1.8 \\
SD &  &  &  & 0.1 & 0.4 & 0.1 & 0.2 \\
\tableline
D & 12208602 & 6 & Medium & -1.8 & -1.5 & -1.9 & -1.3 \\
D & 12208602 & 7 & None reported & -1.4 & -1.2 & -1.5 & -1.0 \\
D & 12208602 & 8 & None reported & -1.0 & -0.8 & -1.2 & -0.8\\
D & 12208602 & 9 & None reported & -1.9 & -1.6 & -2.0 & -1.6 \\
D & 12208602 & 10 & Medium & -2.0 & -1.4 & -2.0 & -1.4 \\
D & 12208602 & 11 & None reported & -1.9 & -1.3 & -1.8 & -1.4 \\
D & 12208602 & 12 & None reported & -1.5 & -1.2 & -1.8 & -0.9\\
D & 12208602 & 13 & None reported & -1.1 & -1.2 & -0.7 & -1.3 \\
\tableline
Mean &  &  &  & -1.5 & -1.2 & -1.5 & -1.2 \\
SD &  &  &  & 0.4 & 0.2 & 0.5 & 0.3 \\
\enddata

\tablenotetext{a}{Means and standard deviations computed using only quarters with no Moir\'e reported, except for A, for which Q6 and Q10 were also included.}

\end{deluxetable}

% -----------------------------------------------------------------
\begin{deluxetable}{cccccc}

\tabletypesize{\scriptsize}

\tablecaption{AGN variability power spectral densities (PSDs) measured with short-cadence Kepler data}

\tablenum{4}

\tablehead{

& & & & \colhead{SAP (raw) data} & \colhead{Corrected SAP data} \\

\colhead{Object} & \colhead{Kepler} &  \colhead{Quarter} & \colhead{Moir\'e} & \colhead{Slope} & \colhead{Slope}  \\

\colhead{Designation} & \colhead{ID} &\colhead{--Month}  &  \colhead{level} & \colhead{(original)} & \colhead{(original \& end-matched)}
} 

%% All data must appear between the \startdata and \enddata commands
\startdata
A & 10663134 & 6--1 & Medium & -2.0 &  -2.0  \\
A & 10663134 & 6--2 & Medium & -2.1 &  -2.0 \\
A & 10663134 & 6--3 & Medium & -1.7 & -1.9 \\
A & 10663134 & 11--1 & High & -1.6 & -2.0  \\
A & 10663134 & 11--2 & High & -2.0 & -2.5 \\
A & 10663134 & 11--3 & High & -1.6  & -1.6  \\
\tableline
Mean\tablenotemark{a}  &  &  &  & -1.9   & -1.9  \\
SD &  &  &  & 0.2 &  0.1  \\
\tableline
B & 11021406 & 6--1 & None reported & -2.0  & -1.9  \\
B & 11021406 & 6--2 & None reported & -1.7  & -1.7  \\
B & 11021406 & 6--3 & None reported & -1.3 &  -1.7  \\
B & 11021406 & 11--1 & None reported & -1.3  & -1.2  \\
B & 11021406 & 11--2 & None reported & -1.3  & -1.3  \\
B & 11021406 & 11--3 & None reported & -1.8  & -1.9  \\
\tableline
Mean &  &  &  & -1.6 &   -1.6   \\
SD &  &  &  & 0.3 &   0.3   \\
\tableline
C & 11606854 & 6--1 & None reported & -1.8  & -2.0  \\
C & 11606854 & 6--2 & None reported & -1.8  & -1.8  \\
C & 11606854 & 6--3 & None reported & -1.9  & -1.8  \\
C & 11606854 & 11--1 & Medium & -1.8  & -2.0  \\
C & 11606854 & 11--2 & Medium & -2.0  & -2.0  \\
C & 11606854 & 11--3 & Medium & -1.9  & -1.8  \\
\tableline
Mean &  &  &  & -1.9 &   -1.9   \\
SD &  &  &  & 0.1 &   0.1   \\
\tableline
D & 12208602 & 6--1 & Medium & -1.9  & -1.7  \\
D & 12208602 & 6--2 & Medium & -1.7  & -1.8  \\
D & 12208602 & 6--3 & Medium & -1.8  & -1.8  \\
D & 12208602 & 10--1 & Medium & -1.8  & -1.9  \\
D & 12208602 & 10--2 & Medium & -1.4 & -1.7  \\
D & 12208602 & 10--3 & Medium & -1.8  & -1.8  \\
\tableline
Mean &  &  &  & -1.7 &   -1.8   \\
SD &  &  &  & 0.2   & 0.1   \\
\enddata

\tablenotetext{a}{Mean and standard deviations computed using only quarters with no or medium Moir\'e reported.}

\end{deluxetable}

\begin{deluxetable}{ccccccccccc}

\tabletypesize{\scriptsize}
\tablewidth{190mm}

\tablecaption{Palomar Observatory Photometry}
\tablenum{5}
\tablehead{\colhead{Object} & \colhead{Filter} & \colhead{Wavelength} & \colhead{Frequency} & \colhead{KIC } & \colhead{Palomar } & \colhead{Error } & \colhead{Reddening} & \colhead{De-reddened} & \colhead{De-reddened} & \colhead{Flux Density} \\ 
\colhead{} & \colhead{Band} & \colhead{} & \colhead{} & \colhead{} & \colhead{} & \colhead{} & \colhead{} &  \colhead{Palomar} & \colhead{Flux Density} & \colhead{Error} \\ 
\colhead{} & \colhead{} & \colhead{(\micron)} & \colhead{(Hz)} & \colhead{(mag)} & \colhead{(mag)} & \colhead{(mag)} & \colhead{(mag)} & \colhead{(mag)} & \colhead{(mJy)} & \colhead{(mJy)} } 
\startdata
% J1923=Object A, 2010-Sep-22, 2010-Dec-05
A\tablenotemark{a} & g' & 0.52 & 5.77E+14 & 18.597 & 19.15 & 0.02 & 0.345 & 18.80 & 0.11 & 0.02 \\
A & r' & 0.67 & 4.48E+14 & 18.591 & 18.62 & 0.02 & 0.244 & 18.37 & 0.20 & 0.02 \\
A & i' & 0.79 & 3.80E+14 & 18.652 & 18.46 & 0.02 & 0.189 & 18.27 & 0.23 & 0.02 \\
A & z' & 0.91 & 3.30E+14 & \nodata\tablenotemark{e} & 18.34 & 0.11 & 0.161 & 18.18 & 0.26 & 0.10 \\
\tableline
% J1909=Object B 2010-Nov-13, 2010-Dec-05, 2010-Dec-09
B\tablenotemark{b} & g' & 0.52 & 5.77E+14 & 18.539 & 18.88 & 0.08 & 0.229 & 18.65 & 0.13 & 0.07 \\
B & r' & 0.67 & 4.48E+14 & 17.973 & 17.99 & 0.05 & 0.168 & 17.82 & 0.34 & 0.05 \\
B & i' & 0.79 & 3.80E+14 & 17.790 & 17.64 & 0.06 & 0.13 & 17.51 & 0.47 & 0.06 \\
B & z' & 0.91 & 3.30E+14 & \nodata\tablenotemark{e} & 17.25 & 0.13 & 0.111 & 17.14 & 0.67 & 0.12 \\
% J1918=Object C, 2011-Nov-12, 2011-Dec-02
\tableline
C\tablenotemark{c} & g' & 0.52 & 5.77E+14 & 17.833 & 17.91 & 0.03 & 0.288 & 17.62 & 0.33 & 0.07 \\
C & r' & 0.67 & 4.48E+14 & 17.723 & 17.77 & 0.01 & 0.199 & 17.57 & 0.34 & 0.05 \\
C & i' & 0.79 & 3.80E+14 & 17.784 & 17.93 & 0.03 & 0.148 & 17.78 & 0.28 & 0.04 \\
C & z' & 0.91 & 3.30E+14 & 17.617 & 18.00 & 0.26 & 0.11 & 17.89 & 0.25 & 0.25 \\
% J1924=Object D, 2011-Nov-12, 2011-Dec-03
\tableline
D\tablenotemark{d} & g' & 0.52 & 5.77E+14 & 18.498 & 18.84 & 0.05 & 0.383 & 18.46 & 0.15 & 0.05 \\
D & r' & 0.67 & 4.48E+14 & 18.394 & 18.23 & 0.05 & 0.282 & 17.95 & 0.30 & 0.04 \\
D & i' & 0.79 & 3.80E+14 & 18.408 & 18.15 & 0.04 & 0.215 & 17.94 & 0.32 & 0.04 \\
D & z' & 0.91 & 3.30E+14 & \nodata\tablenotemark{e} & 18.35 & 0.29 & 0.185 & 18.17 & 0.26 & 0.27 \\
\enddata
\tablenotetext{a}{Observed on 2010-Sep-22, 2010-Dec-05.}
\tablenotetext{b}{Observed on 2010-Nov-13, 2010-Dec-05, 2010-Dec-09.}
\tablenotetext{c}{Observed on 2011-Nov-12, 2011-Dec-02.}
\tablenotetext{d}{Observed on 2011-Nov-12, 2011-Dec-03.}
\tablenotetext{e}{No $z'$ data were available in the KIC for these objects.}

\end{deluxetable}

% -----------------------------------------------------------------
\begin{figure}[ht]
\vspace{0mm}
%\includegraphics[scale=0.80, angle =0]{f1.eps}
% replace f1.eps (7.3 Mbytes) with smaller file for arxiv version and reduce scale to 0.70
\includegraphics[scale=0.70, angle =0]{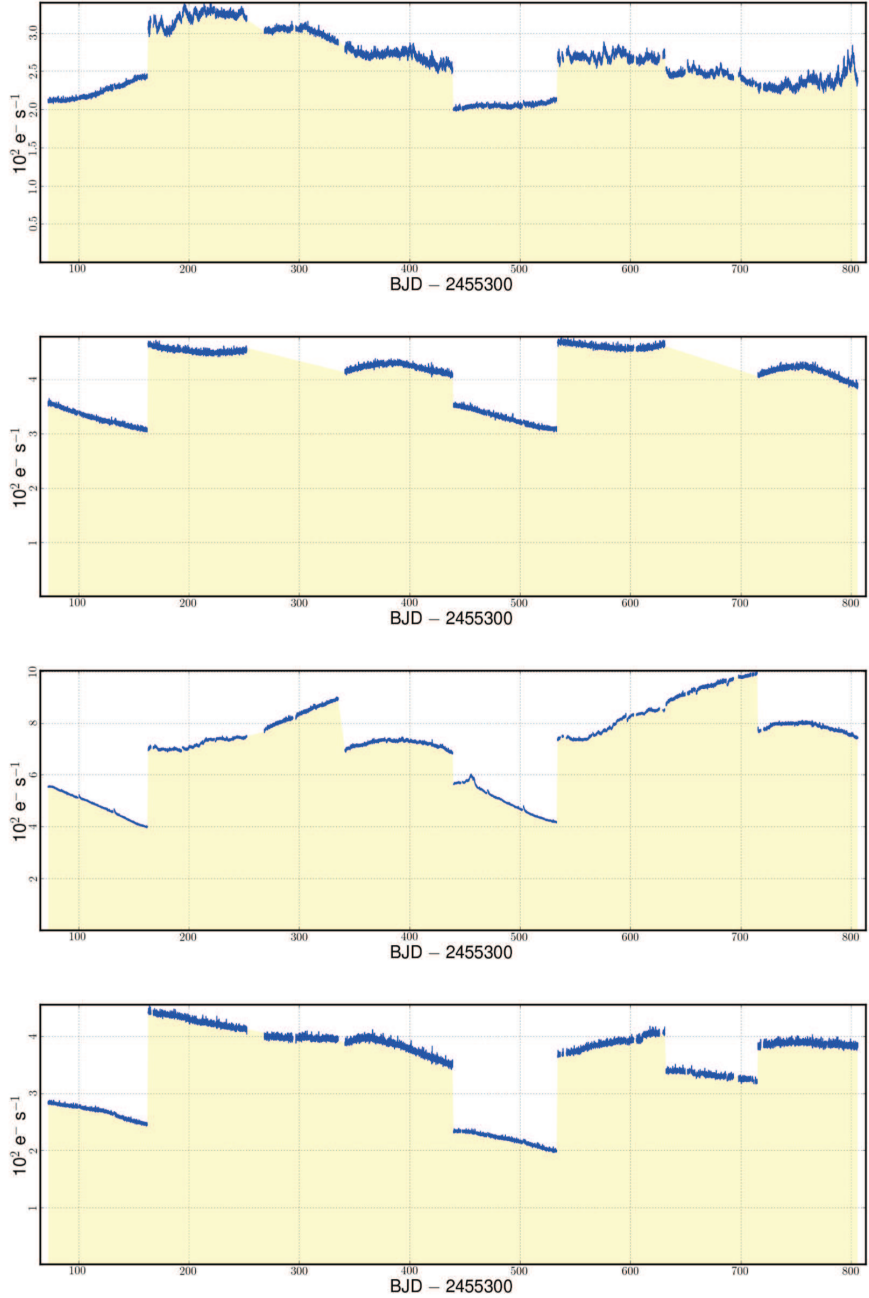}
\caption{Kepler light curves for objects A-D (starting at top) covering Quarters 6-13  in 2010-2012. The vertical axes have units of electrons per second, the horizontal axes have units of days where BJD is the Barycentric Julian Date. See \S 4.2 for an explanation of the steps between quarters.}

\end{figure}
% -----------------------------------------------------------------

\begin{figure}[ht]
\vspace{0mm}
\includegraphics[scale=0.8, angle =0]{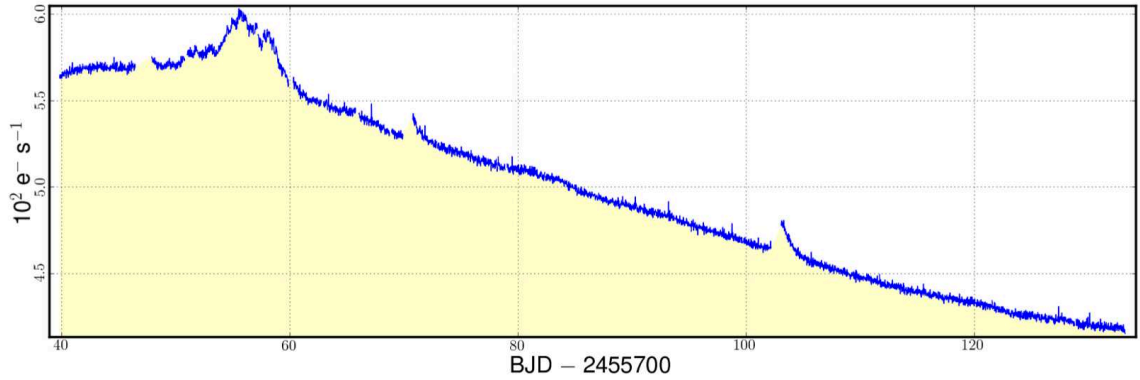}
\caption{Kepler light curve for Object C in Quarter 10. The rise in the amplitude around BJD 2455755 lasting five to six days is clearly visible, as are the two thermal excursions after the monthly data downloads. The vertical axes have units of electrons per second, the horizontal axes have units of days where BJD is the Barycentric Julian Date.}
\end{figure}
% -----------------------------------------------------------------

\begin{figure}[ht]
\vspace{0mm}
\plottwo{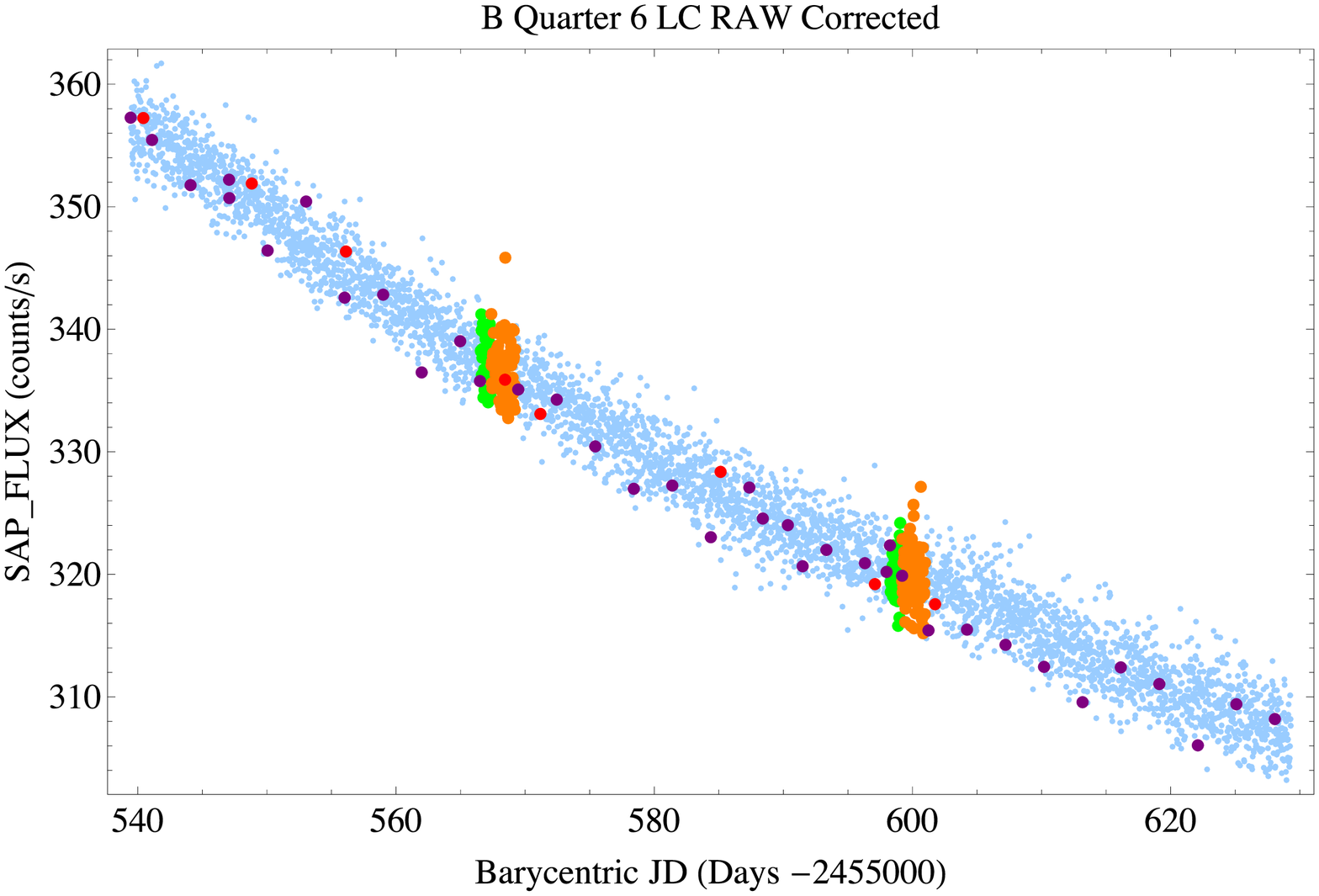}{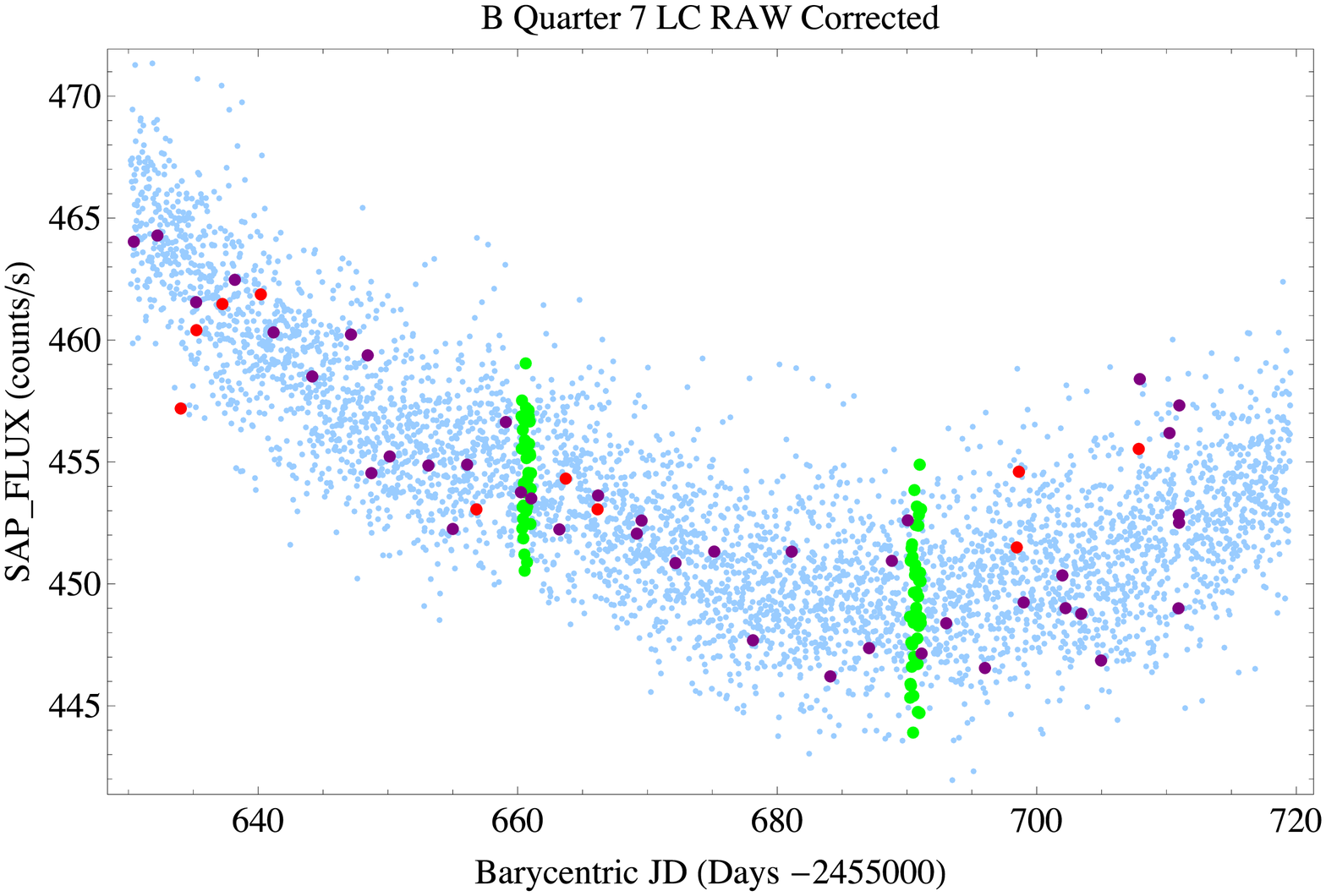}
\vskip 5mm

%\plottwo{images/B-Q9.pdf}{images/B-Q10.pdf}
\plottwo{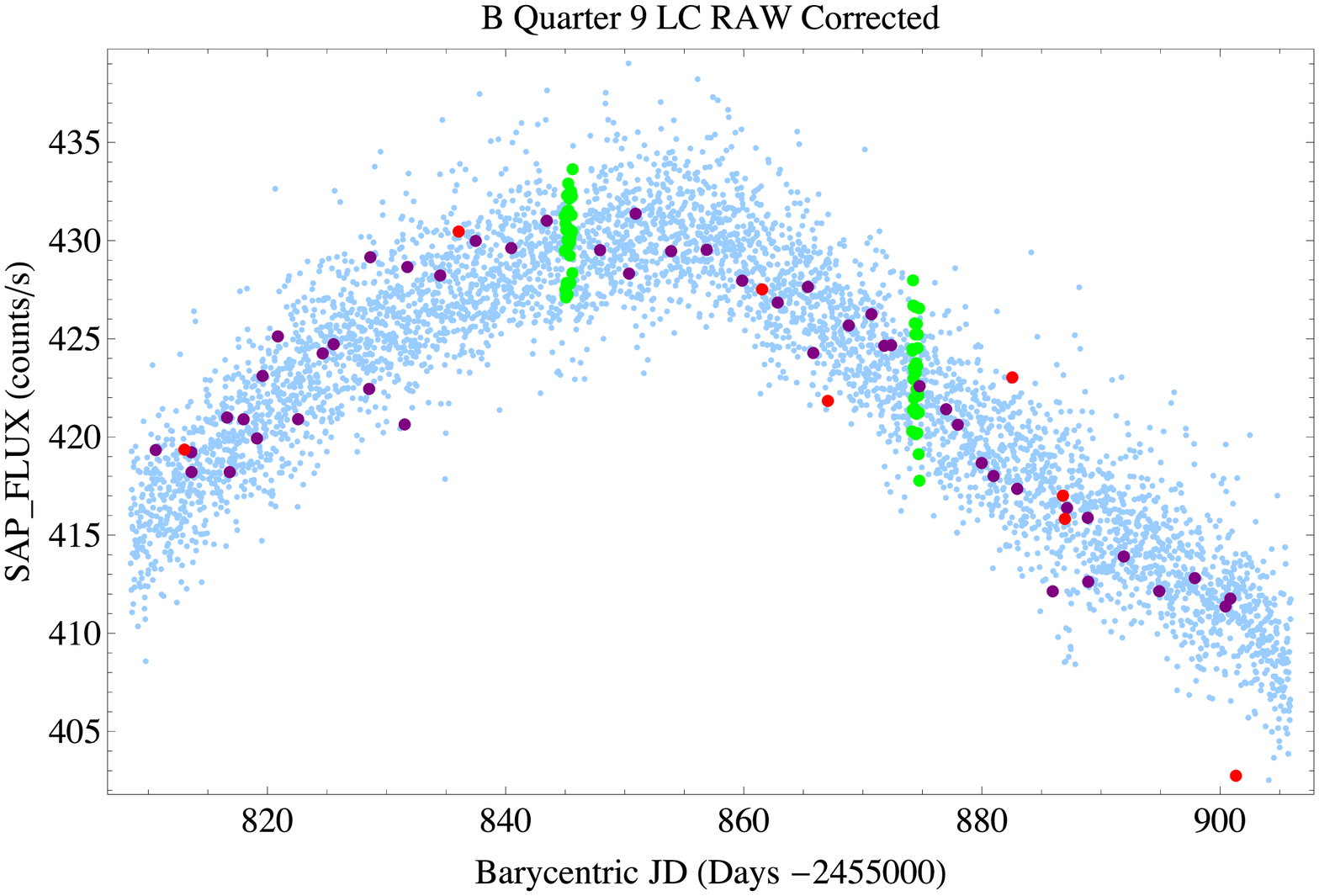}{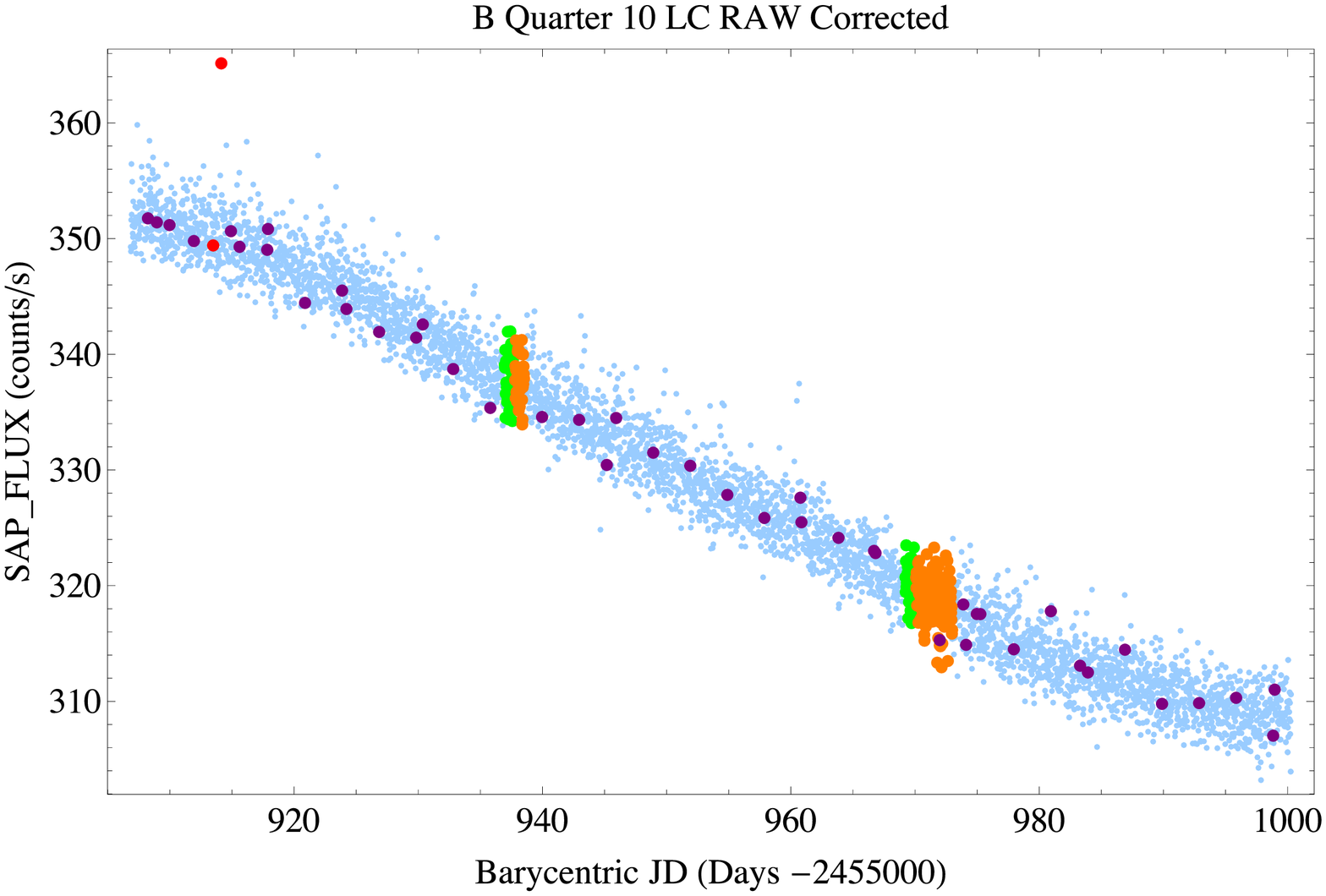}
\vskip 5mm

%\plottwo{images/B-Q11.pdf}{images/B-Q13.pdf}
\plottwo{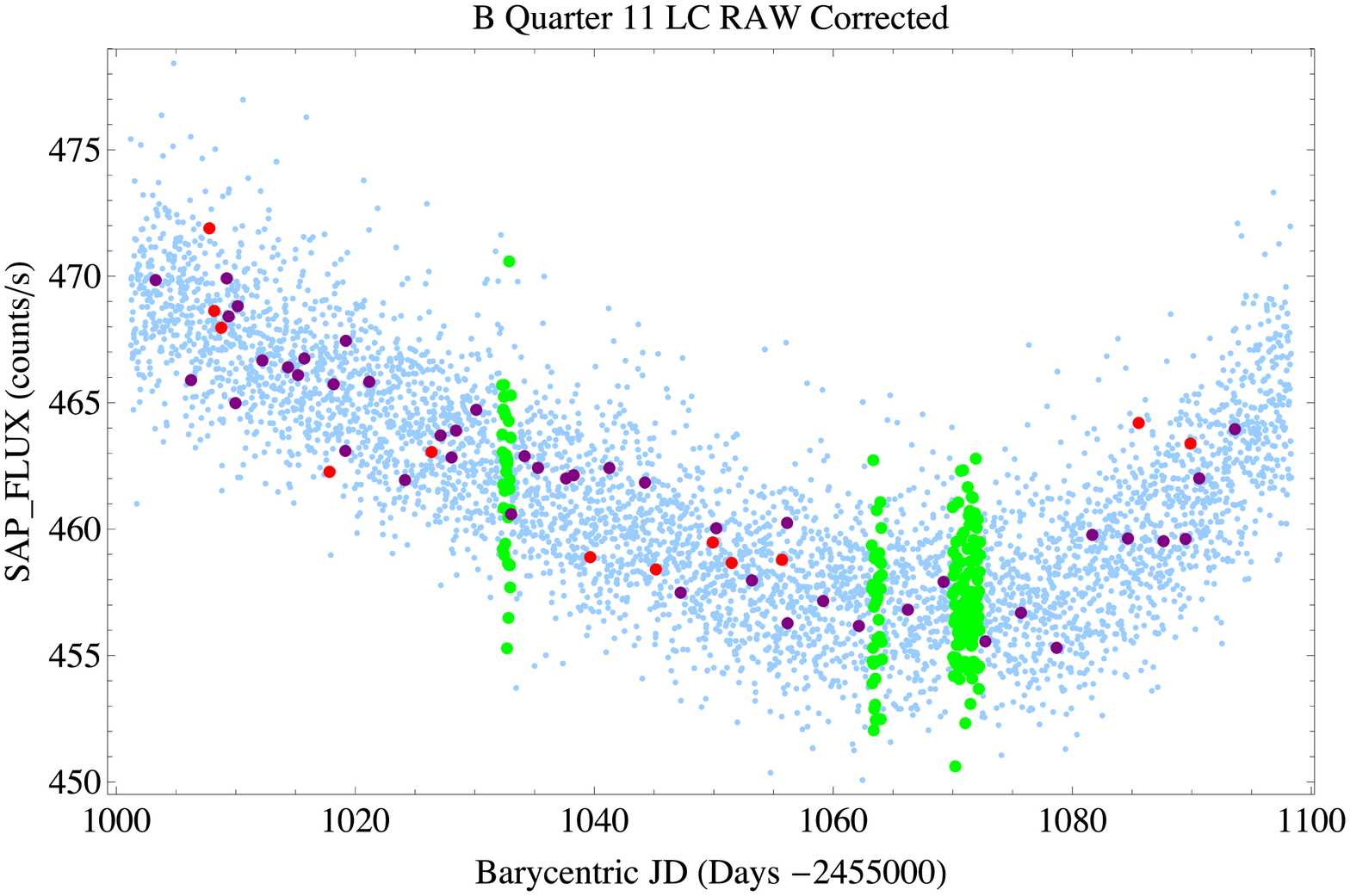}{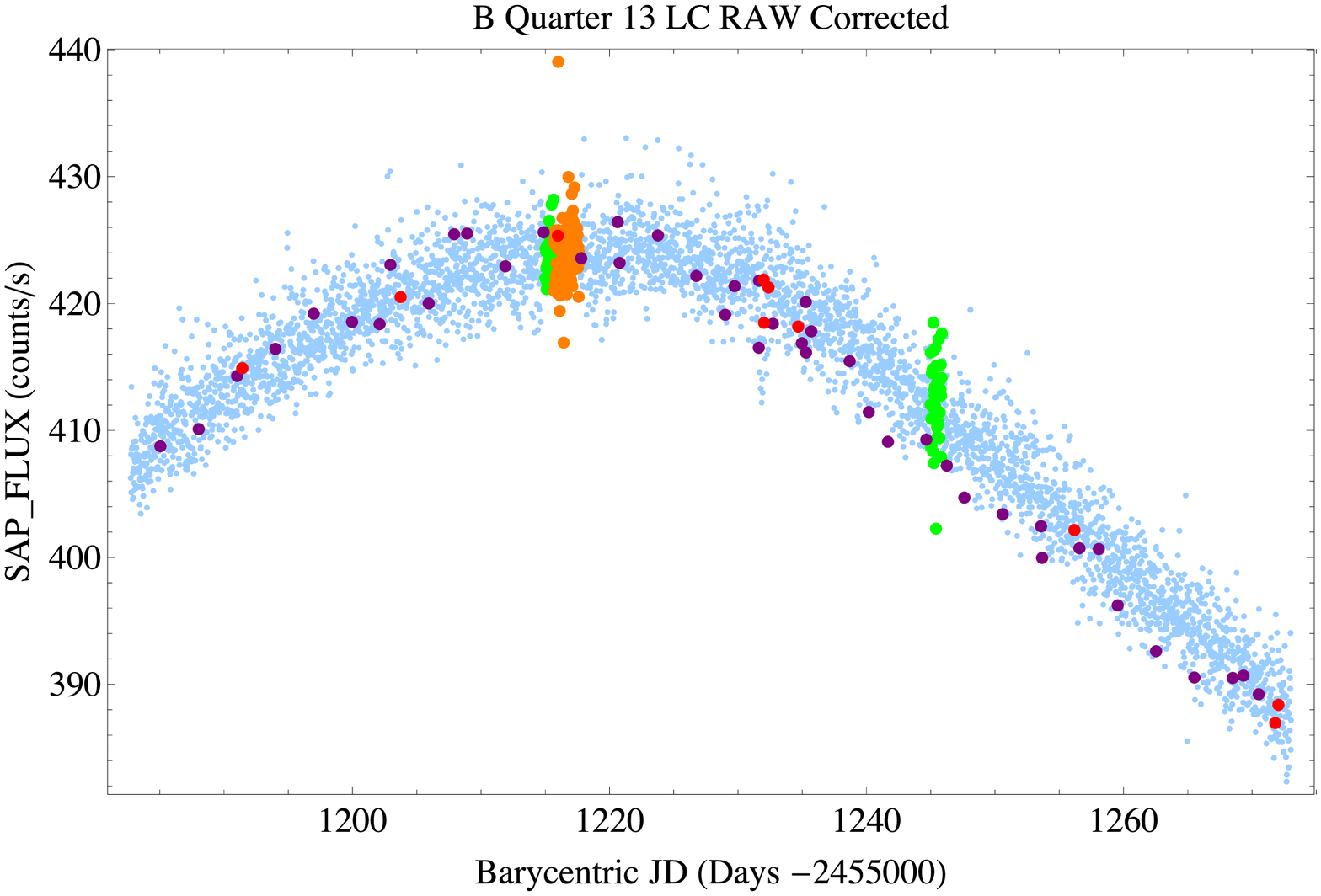}

\caption{Kepler light curves for Object B for 6 quarters (no data were taken for Q8 and Q12 due to an inoperative detector).  Several operations were performed on the ``raw'', or SAP, data prior to computing PSDs (see \S 5). Color coding represents: unmodified data (light blue); removal of monthly thermal glitch (orange); monthly data gap and other missing data (green); data entries with a time value but no flux value (purple); and outliers (red).  The dominant variation is not intrinsic variability, but residual 1-year period  differential velocity aberration and thermal drift instrumental effects \citep{Kinemuchi12}.
}
\end{figure}

% -----------------------------------------------------------------

\begin{figure}[ht]
\vspace{0mm}
%\vskip -10mm
\plottwo{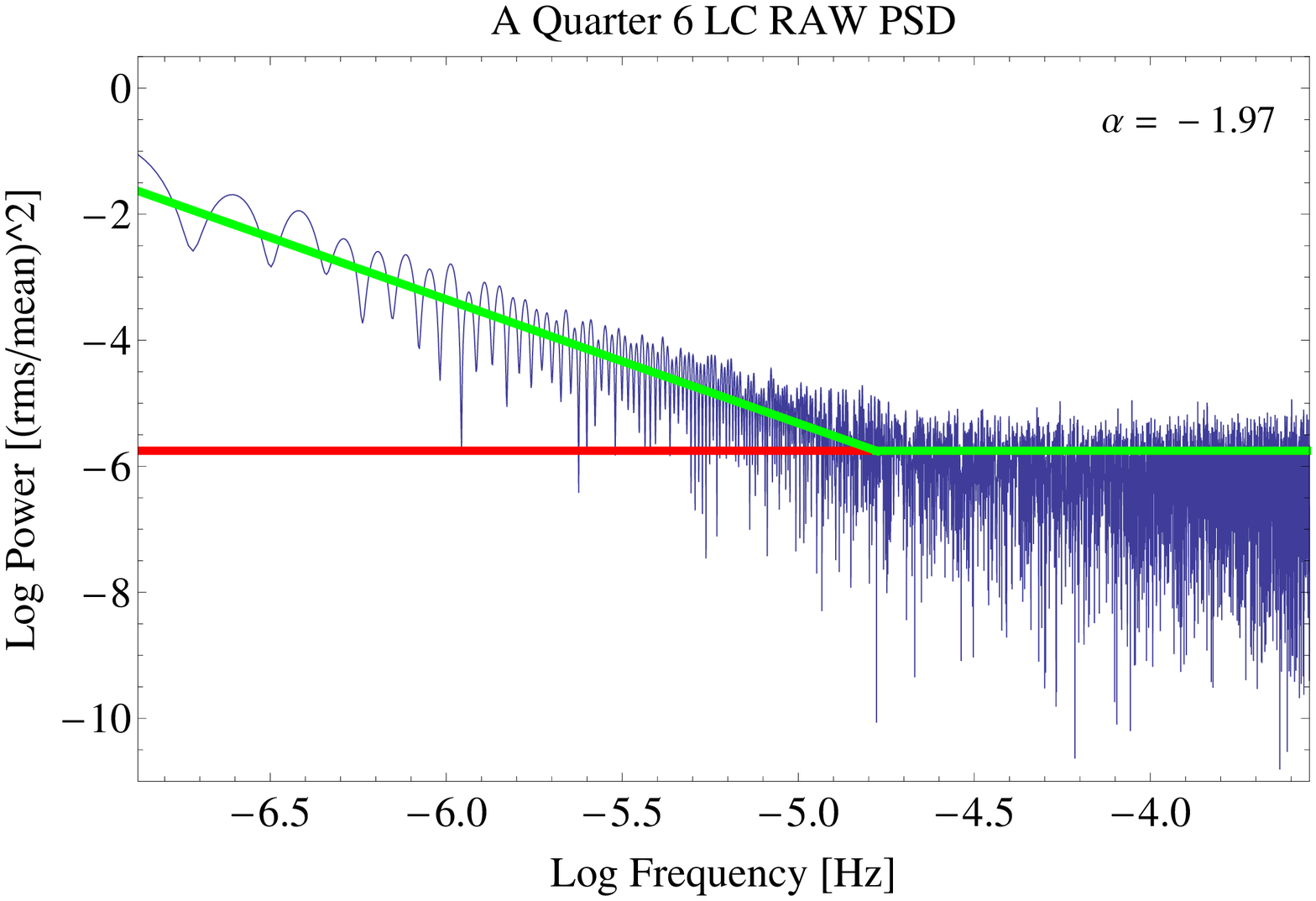}{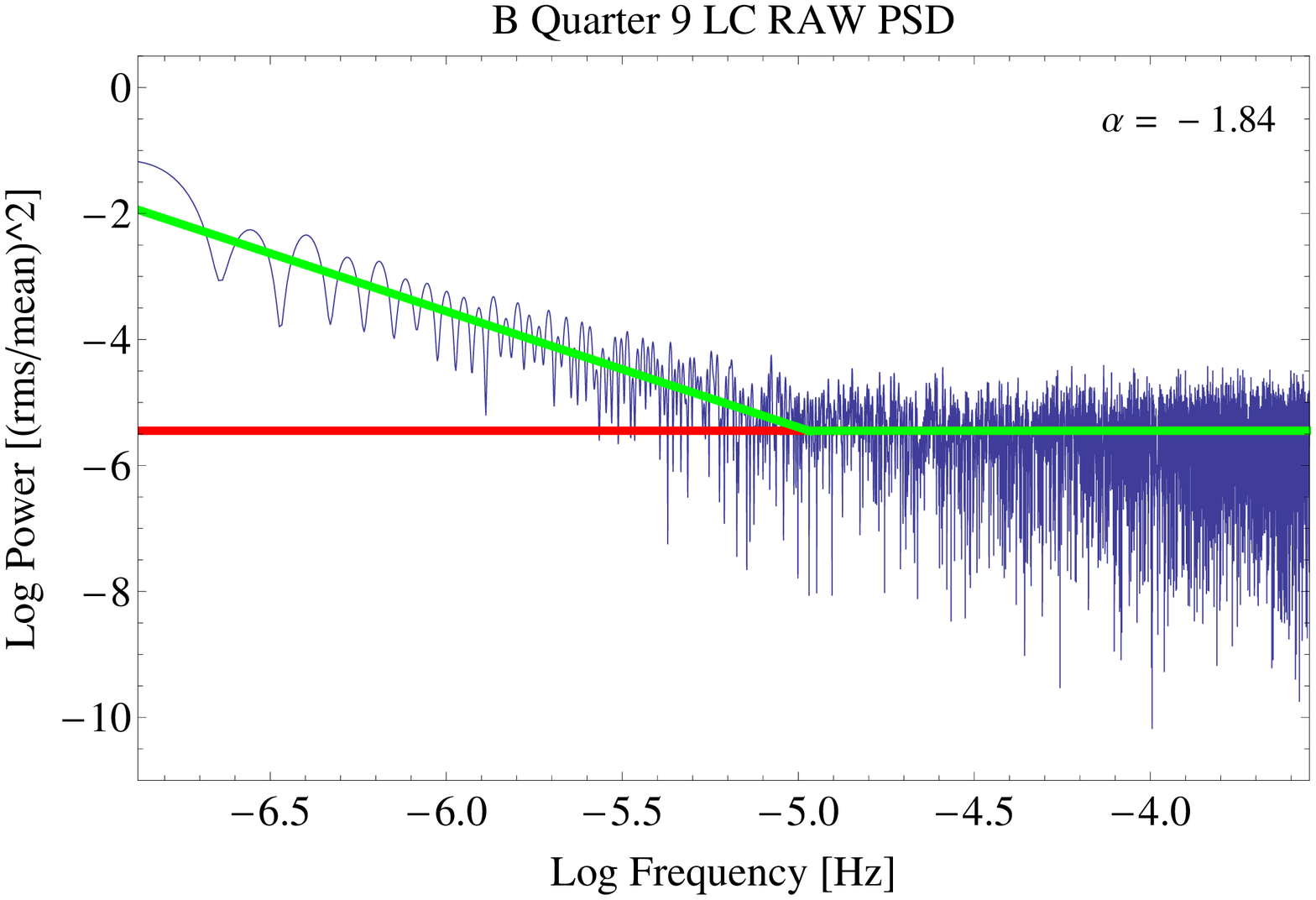}
\vskip 5mm

\plottwo{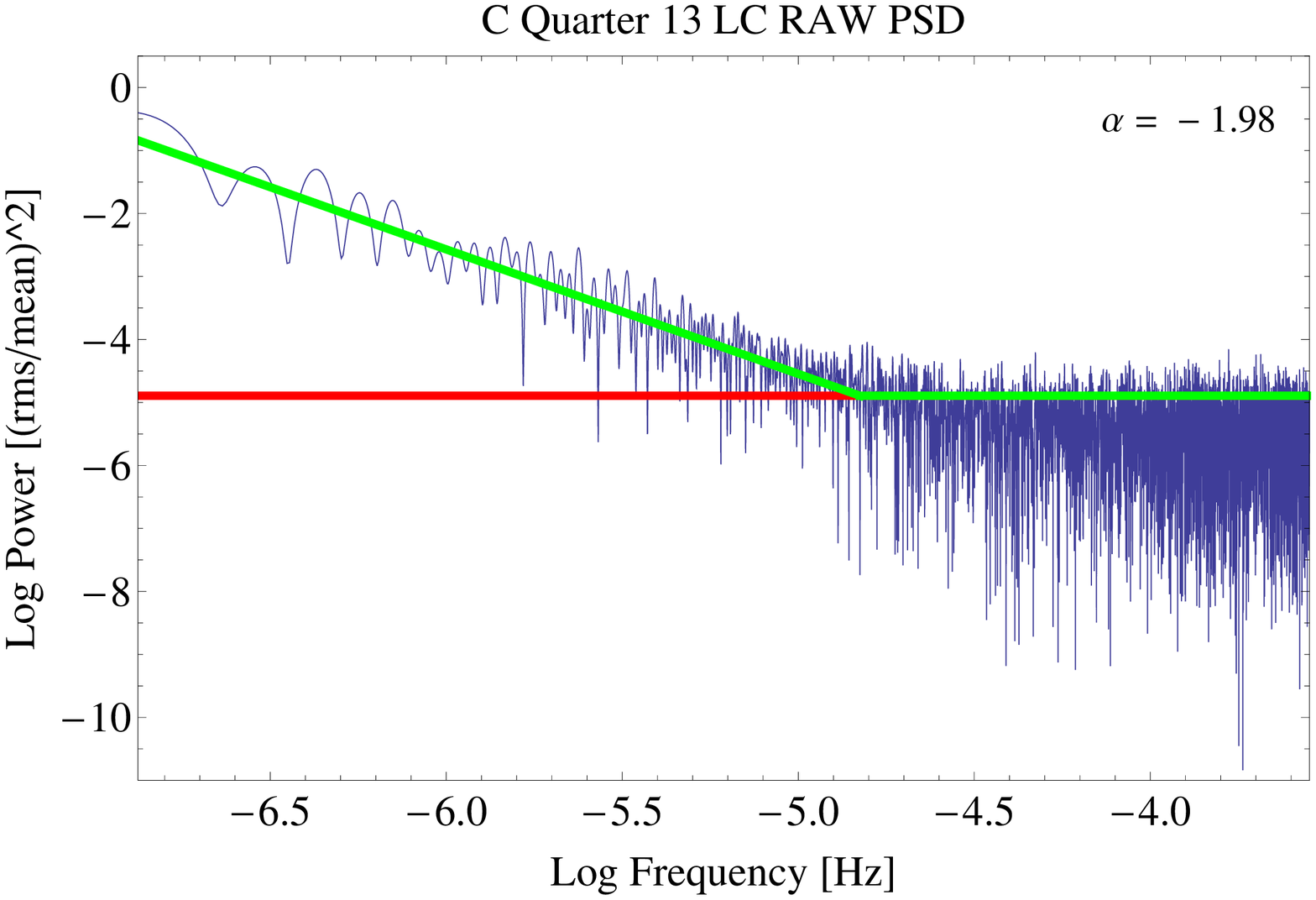}{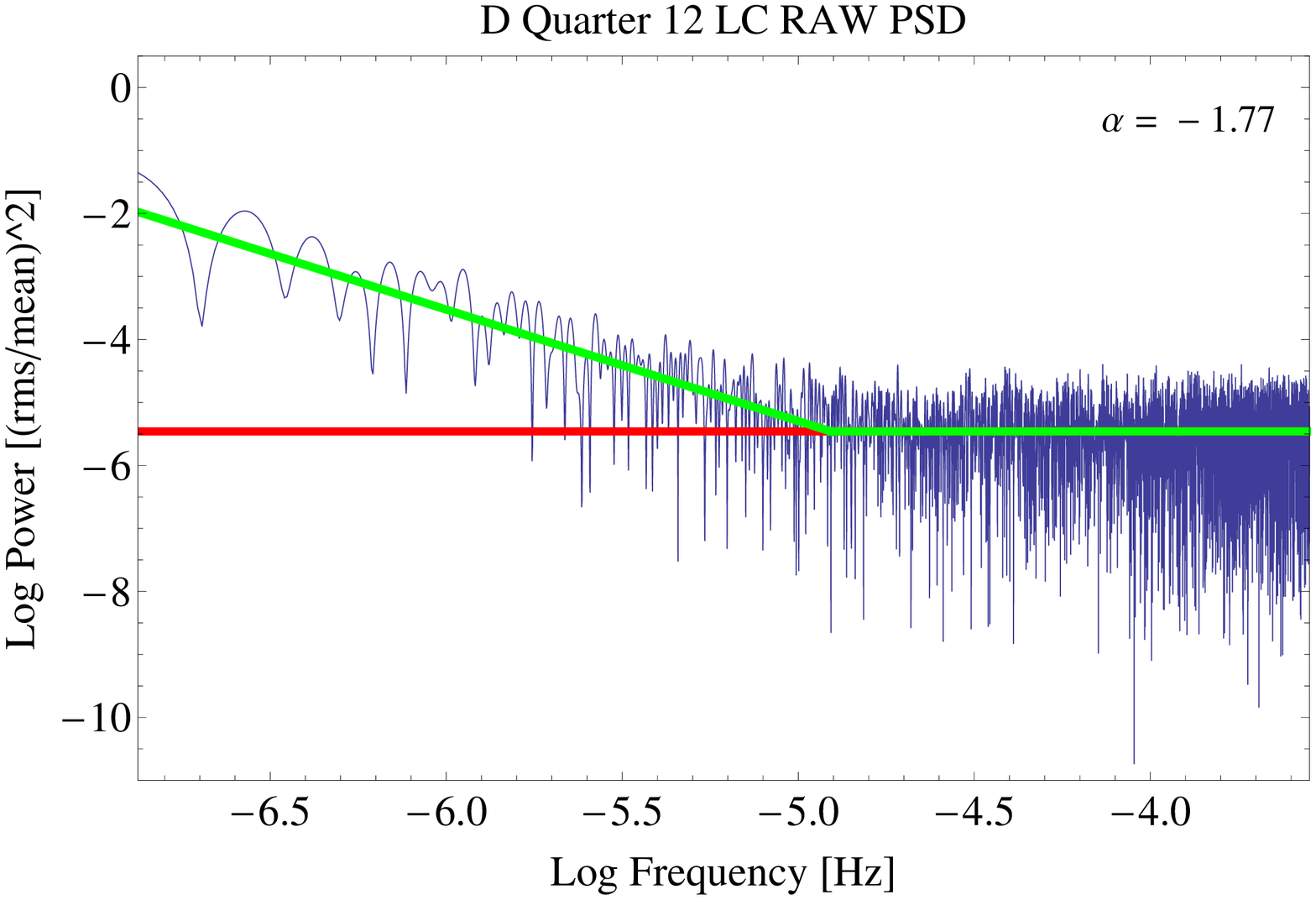}

\caption{Examples of flux variability power spectral densities (PSDs) for Objects A, B, C, and D.   Each shows data from a single quarter, with fluxes corrected for known calibration issues as described in the text.  A two-parameter fit is shown for each: (i) a white-noise level that dominates at high frequencies; and (ii) a power law slope that dominates at frequencies below $\sim 10^{-5}$ Hz.}
\end{figure}


\begin{thebibliography}{}

\bibitem[Bachev et al. (2012)]{bachev12}
Bachev, R. et al. 2012, \mnras, 424, 2625

\bibitem[Bauer et al.(2009)]{bauer2009}
Bauer, A. et al. 2009, \apj, 699, 1732

\bibitem[Borucki et al.(2010)]{borucki2010}
Borucki, W.~J. et al. 2010, Science, 327, 977

\bibitem[Brown et al.(2011)]{brown11}
Brown, T. M., Latham, D. W., Everett, M. E., and Esquerdo, G. A. 2011, \aj, 142, 112

\bibitem[Britzen et al.(2008)]{britzen2008}
Britzen, S. et al. 2008, \aap, 484, 119

\bibitem[Bryson et al.(2010)]{bryson2010}
Bryson, S.~T. et al. \apj, 713, L97

\bibitem[Carini et al.(2007)]{carini07}
Carini, M.~T., Noble, J.~C., Taylor, R., \& Culler, R.  2007, \aj, 133, 303

\bibitem[Carini \& Ryle(2012)]{cr12}
Carini, M.~T., \& Ryle, W.~T. 2012, \apj, 749, 70

\bibitem[Collier \& Peterson(2001)]{coll01}
Collier, S., \& Peterson, B. M. 2001, ApJ, 555, 775

\bibitem[de Vries et al.(2005)]{devries05}
de Vries, W. H., Becker, R. H., White, R. L., \& Loomis, C. 2005, AJ, 129, 615

\bibitem[Edelson \& Nandra(1999)]{ede99}
Edelson, R., \& Nandra, K. 1999, \apj, 514, 682

\bibitem[Fougere(1985)]{fou85}
Fougere, P.~F. 1985, J. Geophys. Res., A5, 4355

\bibitem[Fukugita et al.(1996)]{fukugita1996}
Fukugita, M. et al. 1996, \aj, 111, 1748

\bibitem[Giannios, Uzdensky \& Begelman(2009)]{giannos09}
Giannos, D., Uzdensky, D.~A, \& Begelman, M. 2009, \mnras, 395, L29

\bibitem[Giannios, Uzdensky \& Begelman(2010)]{giannios10}
Giannos, D., Uzdensky, D.~A, \& Begelman, M. 2010, \mnras, 402, 1649

\bibitem[Gierli{\'n}ski et al.(2008)]{Gierlinski08}
Gierli{\'n}ski, M., Middleton, M., Ward, M. \&  Done, C.  2008, \nat, 455, 369

\bibitem[Gopal-Krishna et al.(2003)]{gk03}
Gopal-Krishna, Stalin, C.~S., Sagar, R. \& Wiita, P.~J. 2003, \apj, 586, L25

\bibitem[Gopal-Krishna et al.(2011)]{gk11}
Gopal-Krishna, Goyal, A., Joshi, S., Karthick, C., Sagar, R., Wiita, P.~J., Anupama, G.~C., \& Sahu, D.~K. 2011, \mnras, 416, 101

\bibitem[Goyal et al.(2012)]{goyal12}
Goyal, A., Gopal-Krishna, Wiita, P.~J., Anupama, G.~C., Sahu, K.~D., Sagar, R., \& Joshi, S.  2012, \aap, 544, 37

\bibitem[Gupta, Srivastava \& Wiita(2009)]{gsw09}
Gupta, A.~C., Srivastava, A.~K. \& Wiita, P.~J. 2009, \apj, 690, 216

\bibitem[Hawkins(2002)]{hawk02}
Hawkins, M.~R.~S. 2002, \mnras, 329, 76

\bibitem[Henstock et al.(1997)]{hen97}
Henstock, D. R., Browne, I.W.A., Wilkinson, P.N, \& McMahon, R. G. 1997, \mnras, 290, 380.

\bibitem[Hughes et al.(1992)]{hughes92}
Hughes, P. A., Aller, H. D., \& Aller, M. F. 1992, ApJ, 396, 469

\bibitem[Healey et al.(2008)]{healey2008}
Healey, S. et al. 2008, \apjs, 175, 97

\bibitem[Jenkins et al.(2010)]{jenkins2010}
Jenkins, J. M. et al. 2010 \apj, 713, L87

\bibitem[Jorstad et al.(2005)]{jorstad05}
Jorstad, S., et al. 2005, \aj, 130, 1418

\bibitem[Kartaltepe \& Balonek(2007)]{Kartaltepe07}
Kartaltepe, J.~S. \& Balonek, T.~J. 2007, \aj, 133, 2866

\bibitem[Kepler Project(2009a)]{kepler2009a}
Kepler Project, 2009 "Kepler Instrument Handbook" KSCI-19033-001, (Moffet Field, CA: NASA Ames Research Center), 47

\bibitem[Kepler Project(2009b)]{kepler2009b}
Kepler Project, 2009, "Kepler Archive Manual" KDMC-10008, (Moffet Field, CA: NASA Ames Research Center)

\bibitem[Kepler Project(2012)]{kepler2012}
Kepler Project, 2012 "Kepler Data Release 17 Notes" KSCI-19057-001, (Moffet Field, CA: NASA Ames Research Center)

\bibitem[Kinemuchi et al.(2012)]{Kinemuchi12} 
Kinemuchi, K., et al.\ 2012, \pasp, 124, 963 


\bibitem[Koch et al.(2010)]{koch2010}
Koch, D.~G. et al. 2010, \apj, 713, L29

\bibitem[Kolodziejczak et al.(2010)]{Kolodziejczak10} 
Kolodziejczak, J.~J., Caldwell, D.~A., van Cleve, J.~E., et al.\ 2010, \procspie, 7742,  77421G

\bibitem[Lachowicz et al.(2009)]{lac09}
Lachowicz, P., Gupta, A.~C., Gaur, H. \& Wiita, P.~J. 2009, A\&A, 506, L17

\bibitem[Lenz \& Breger(2009)]{lenz05}
Lenz, P., \& Breger, M. 2005, CoAst, 146, 53

\bibitem[MacLeod et al.(2010)]{macleod10}
MacLeod, C. et al. 2010, \apj, 721, 1014

\bibitem[MacLeod et al.(2012)]{macleod12}
MacLeod, C. et al. 2012, \apj, 753, 106

\bibitem[Mangalam \& Wiita(1993)]{mw93}
Mangalam, A.~V. \& Wiita, P.~J. 1993, \apj, 406, 420

\bibitem[Markowitz et al.(2003)]{markowitz03}
Markowitz, A., et al. 2003, \apj, 593, 96

\bibitem[Markowitz(2009)]{mar09}
Markowitz, A. 2009, \apj, 698, 1740

\bibitem[Marscher \& Gear(1985)]{mg85}
Marscher, A.~P., \& Gear, W.~K. 1985, \apj, 298, 114

\bibitem[Marscher \& Travis(1991)]{mt91}
Marscher, A.~P., \& Travis, J.~P. 1991, in Variability of Active Galactic Nuclei, eds.\ H.~R. Miller \& P.~J. Wiita (Cambridge: Cambridge U.~P.), p.\ 153

\bibitem[Marscher et al.(2008)]{marscher08}
Marscher, A.~P., et al. 2008, \nat, 452, 966

\bibitem[Miller(1996)]{miller96}
Miller, H.~R. 1996, \pasp, 110, 17

\bibitem[Moran(1996)]{moran1996}
Moran, E. C., Helfand, D., Becker, R. H., and White, R. L. 1996, \apj, 461, 127.

\bibitem[Mushotzky et al.(2011)]{mushotzky11}
Mushotzky, R., Edelson, R., Baumgartner, W., \& Gandhi, P.  2011, \apj, 743, L12

\bibitem[Nalewajko et al.(2011)]{nalewajko11}
Nalewajko, K. Giannos, D. , Begelman, M.~C., Uzdensky, D. A. \& Sikova, M. 2011, \mnras, 413, 333

\bibitem[Osterman Meyer et al.(2009)]{osterman09}
Osterman Meyer, A., et al. 2009, \aj, 138, 1902

\bibitem[Pica et al.(1988)]{pica88}
Pica, A.~J., Smith, A.~G., Webb, J.~R., Leacock, R~.J., Clements, S., \& Gombola, P.~P. 1988, \aj, 96, 1215

\bibitem[Quirrenbach et al.(1991)]{quirrenbach91}
Quirrenbach, A. et al. 1991, \apj, 372, L71

\bibitem[Ram{\'i}rez et al.(2009)]{ramirez09}
Ram{\'i}rez, A., de Diego, J. A., Dultzin, D., \& Gonz{\'a}lez-P{\'e}rez, J.-N. 2009, \aj, 138, 991

\bibitem[Rani et al.(2009)]{rani09}
Rani, B., Wiita, P.~J. \& Gupta, A.~C.  2009, \apj, 696, 2170

\bibitem[Rani et al.(2010)]{rani10}
Rani, B., Gupta, A.~C., Joshi, U.~C., Ganesh, S., \& Wiita, P.~J. 2010, \apj, 719, L153

\bibitem[Ruan et al.(2012)]{ruan2012}
Ruan, J.~J. et al. 2012, \apj, 760, 51

\bibitem[Sagar et al.(2004)]{sagar04}
Sagar, R., Stalin, C.~S., Gopal-Krishna, \& Wiita, P.~J. 2004, \mnras, 348, 176

\bibitem[Stalin et al.(2004)]{stalin04}
Stalin, C.~S., Gopal-Krishna, Sagar, R., \& Wiita, P.~J. 2004, JApA, 25, 1 

\bibitem[Still and Barclay(2012)]{still2012}
Still, M. \& Barclay, T. 2012, Astrophysics Code Library, record ascl: 1208.004. ``PyKe" contributed software available via download from http://keplergo.arc.nasa.gov/PyKE.shtml 

\bibitem[Sun \& Malkan(1989)]{sm89}
Sun W.-H., \& Malkan, M. A. 1989, \apj, 346, 68

\bibitem[Taylor(2005)]{taylor2005}
Taylor, M. 2005, ASP Conference Series 347, 29.  ``topcat" software available via download from http://www.star.bris.ac.uk/~mbt/topcat/ 

\bibitem[Urry \& Padovani(1995)]{up95}
Urry, C.~M. \& Padovani, P. 1995, \pasp, 107, 803

\bibitem[Vanden Berk et al.(2004)]{vanden04}
Vanden Berk, D. E., et al. 2004, \apj, 601, 692

\bibitem[Vaughn et al.(2003)]{vaughan2003}
Vaughan, S. et al. 2003, \mnras, 345, 1271

\bibitem[York et al.(2000)]{york00}
York, D.~G. et al. 2000, \aj, 120, 1579

\bibitem[Zrake and MacFadyen(2013)]{zrake2013}
Zrake, J. \& MacFadyen, A. I. 2013 \apj, 763, L12

\end{thebibliography}
\end{document}